\documentclass[%
 aip,
 amsmath,amssymb,
 reprint,%
]{revtex4-1}

\usepackage{graphicx}
\usepackage{dcolumn}
\usepackage{bm}
\usepackage[utf8]{inputenc}
\usepackage[T1]{fontenc}
\usepackage{mathptmx}
\usepackage{etoolbox}
\usepackage{float}
\usepackage{fancyhdr}
\usepackage{fnpos}
\usepackage[english]{babel}
\usepackage{array}
\usepackage{droidsans}
\usepackage{charter}
\usepackage[T1]{fontenc}
\usepackage[usenames,dvipsnames]{xcolor}
\usepackage{setspace}
\usepackage[compact]{titlesec}
\usepackage{hyperref}
\usepackage{mathptmx}
\usepackage{scalerel,amsmath}
\usepackage{xcolor}
\usepackage{braket}
\usepackage{adjustbox}
\usepackage{makecell}
\usepackage{soul}

\makeatletter
\def\@email#1#2{%
 \endgroup
 \patchcmd{\titleblock@produce}
  {\frontmatter@RRAPformat}
  {\frontmatter@RRAPformat{\produce@RRAP{*#1\href{mailto:#2}{#2}}}\frontmatter@RRAPformat}
  {}{}
}%
\makeatother
\begin{document}

\preprint{AIP/123-QED}

\title
[Machine-Learning-Based Construction of Molecular Potential ...
% and Its Application in Exploring the Deep-Lying-Orbital Effect in High-Order Harmonic Generation
]
{Machine-Learning-Based Construction of Molecular Potential and \\
Its Application in Exploring the Deep-Lying-Orbital Effect in \\
High-Order Harmonic Generation
}

\author{Duong D. Hoang-Trong}
    \affiliation{Computational Physics Key Laboratory K002, Department of Physics, Ho Chi Minh City University of Education, 280 An Duong Vuong Street, Ward 4, District 5, Ho Chi Minh City 72711, Vietnam}

\author{Khang Tran}
    \affiliation{Department of Informatics, New Jersey Institute of Technology, Newark, NJ 07102, USA}

\author{Doan-An Trieu}
    \affiliation{Computational Physics Key Laboratory K002, Department of Physics, Ho Chi Minh City University of Education, 280 An Duong Vuong Street, Ward 4, District 5, Ho Chi Minh City 72711, Vietnam}

\author{Quan-Hao Truong}
    \affiliation{Computational Physics Key Laboratory K002, Department of Physics, Ho Chi Minh City University of Education, 280 An Duong Vuong Street, Ward 4, District 5, Ho Chi Minh City 72711, Vietnam}

\author{Ngoc-Hung Phan}
    \affiliation{Computational Physics Key Laboratory K002, Department of Physics, Ho Chi Minh City University of Education, 280 An Duong Vuong Street, Ward 4, District 5, Ho Chi Minh City 72711, Vietnam}

\author{Ngoc-Loan Phan} 
    \affiliation{Computational Physics Key Laboratory K002, Department of Physics, Ho Chi Minh City University of Education, 280 An Duong Vuong Street, Ward 4, District 5, Ho Chi Minh City 72711, Vietnam}

\author{Van-Hoang Le}
    % \email{hoanglv@hcmue.edu.vn.}  
    \affiliation{Computational Physics Key Laboratory K002, Department of Physics, Ho Chi Minh City University of Education, 280 An Duong Vuong Street, Ward 4, District 5, Ho Chi Minh City 72711, Vietnam}
    
\date{\today}% It is always \today, today,

\begin{abstract}
Creating soft-Coulomb-type (SC) molecular potential within single-active-electron approximation (SAE) is essential since it allows solving time-dependent Schr{\"o}dinger equations with fewer computational resources compared to other multielectron methods. The current available SC potentials can accurately reproduce the energy of the highest occupied molecular orbital (HOMO), which is sufficient for analyzing nonlinear effects in laser-molecule interactions like high-order harmonic generation (HHG). 
However, recent discoveries of significant effects of deep-lying molecular orbitals call for more precise potentials to analyze them. In this study, we present a fast and accurate method based on machine learning to construct SC potentials that simultaneously reproduce various molecular features, including energies, symmetries, and dipole moments of HOMO, HOMO-1, and HOMO-2. We use this ML model to create SC SAE potentials of the HCN molecule and then comprehensively analyze the fingerprints of lower-lying orbitals in HHG spectra emitted during the H-CN stretching.
Our findings reveal that HOMO-1 plays a role in forming the second HHG plateau. Additionally, as the H-C distance increases, the plateau structure and the smoothness of HHG spectra are altered due to the redistribution of orbital electron density. These results are in line with other experimental and theoretical studies. Lastly, the machine learning approach using deconvolution and convolution neural networks in the present study is so general that it can be applied to construct molecular potential for other molecules and molecular dynamic processes.
\end{abstract}

\maketitle

% \begin{quotation}
% The ``lead paragraph'' is encapsulated with the \LaTeX\ 
% \verb+quotation+ environment and is formatted as a single paragraph before the first section heading. 
% (The \verb+quotation+ environment reverts to its usual meaning after the first sectioning command.) 
% Note that numbered references are allowed in the lead paragraph.
% %
% The lead paragraph will only be found in an article being prepared for the journal \textit{Chaos}.
% \end{quotation}

%%%% Start %%%%%%
\section{INTRODUCTION} 
\label{Sec1} 
 Recent decades with advanced laser technologies have witnessed an enduring endeavor of strong-field physics, focusing on the interaction of matter and intense laser field~\cite{Krausz:RevModPhys09,Calegari:jpb16,Li:NatCom20}. It has led to the discovery and celebration of various highly nonlinear phenomena, including high-order harmonic generation (HHG), above-threshold ionization (ATI), high-energy ATI (HATI), and non-sequential double ionization (NSDI)~\cite{Huillier:jpb91,Krause:prl92,Corkum:prl93,Lewenstein:pra94,Lewenstein:pra95}. These nonlinear phenomena have sparked great interest due to their potential applications in generating extraordinary coherent XUV and even soft X-ray radiations, producing single attosecond pulses or attosecond pulse trains, as well as enabling time-resolved imaging and probing of dynamics inside atoms and molecules~\cite{Krausz:RevModPhys09,Calegari:jpb16,Li:NatCom20,Itatani:Nat04,le:PRA07,Haessler:NatPhys10,Kraus:Sci15,He:NatCom22}. Given their promising applications, it is crucial to accurately interpret experimental observations and understand the underlying physics, that necessitates precise theoretical descriptions of these nonlinear phenomena. One common approach is directly solving the time-dependent Schr{\"o}dinger equation (TDSE) of a specific atomic or molecular potential model, ensuring the asymptotic behavior of the wave functions~\cite{Feit:JCP82,Bauer:CPC06,Peters:pra12,Zhu:pra15,Wang:pra20}. The potential model is often parameterized as an analytical expression of the soft-Coulomb-type (SC) potential with one or a few parameters~\cite{Xu:pra09,Peters:pra12, Zhu:pra15,Vu:jpb17,Wang:pra20,Long:jpb23}. Constructing a potential that matches experimental observations or fully quantum \textit{ab initio} simulations is a challenging but not trivial task due to its high coherence and nonlinearity~\cite{Xu:pra09,Abu-samha:pra10,Abu:pra20,Shvetsov:jpb21}. Most atomic or molecular potential models require the energy of the highest occupied molecule orbital (HOMO) to mimic that of a real target, as the HOMO dominantly contributes to highly nonlinear spectra compared to lower-lying orbitals~\cite{Itatani:Nat04,Xu:pra09,Peters:pra12}. This simple matching can be achieved through traditional optimization techniques, as numerous sets of potential parameters yield the same HOMO energy. 

However, recent critical studies have progressively revealed the role of lower-lying molecular orbitals in observed nonlinear optical phenomena~\cite{Mikosch:prl13,Winney:JPCL18,Luo:JPCA17,Liao:cjc16,Ohmura:jpb20}, specifically in HHG~\cite{McFarland:Sci08,Li:Sci08,Worner:prl09,Shiner:NatPhys11,Smirnova:Nat09,Long:jpb23,Yun:prl15,Shu:prl22,Camper:prl23,Chu:pra23,Koushki:jmm23}. There is significant coupling between HOMO and lower-lying orbitals such as \mbox{HOMO-1} or \mbox{HOMO-2}, observable in all three regions of HHG spectra, including the cutoff~\cite{McFarland:Sci08,Li:Sci08,Chu:pra23,Koushki:jmm23}, plateau~\cite{Worner:prl09,Shiner:NatPhys11,Smirnova:Nat09,Chu:pra23}, below- and near-threshold harmonic regions~\cite{Long:jpb23,Chu:pra23}. More recently, it has been discovered that coupling with even lower-lying orbitals, such as HOMO-3 for CO$_2$~\cite{Shu:prl22,Camper:prl23} and N$_2$O~\cite{Fu:oe23}, and even with HOMO-4 for uracil~\cite{Luppi:JCPA23}, can lead to novel structures in HHG spectra. It is important to emphasize that the multiorbital coupling shown in HHG provides insight into the electron-electron interaction encoding dynamics in the attosecond time scale, which are otherwise challenging to observe~\cite{Li:Sci08,Haessler:NatPhys10,Kraus:Sci15,He:NatCom22,Camper:prl23,Peng:NatRevPhys19}. 

In order to theoretically explore these effects of deep-lying orbitals, a necessary prerequisite is a reliable potential model which accurately reproduces features of not only HOMO but also lower-lying orbitals. Previous studies~\cite{Xu:pra09,Shvetsov:jpb21} have employed traditional optimization procedures to determine atomic potential (from the ATI or HATI spectra) that gives accurate energies of the bound ground and a few excited states. However, this method is limited to atomic targets. The reconstruction of the electronic potential of molecules faces difficulties due to the growing number of potential parameters that need to be optimized. Although the SC single-active electron (SAE) potential for nitrogen molecules has been constructed for HOMO and HOMO-1 orbitals, their parameter sets are distinct for each orbital~\cite{Long:jpb23}. The challenge becomes more complex when working with polyatomic molecules containing more than two atoms, as the molecular potential must accommodate a reasonable electron arrangement for each orbital. 

On the other hand, the problem is more complicated for polar molecules since both the molecular polarization and the multichannel coherence are involved in orientation-dependent~\cite{Zhang:pra14,Le:pra18,Koushki:jmm23} and odd-even patterns~\cite{Kraus:prl14,Nguyen:pra22,Chu:pra23} of HHG. In these cases, the potential model also needs to account for polarizations in addition to energies and symmetries of orbitals within real polar molecules and accurately depict the asymptotic behavior of the wave functions. However, constructing a molecular potential that simultaneously meets these multiple constraints is nontrivial due to its highly nonlinear and coherent nature, making applying traditional statistical learning methods challenging.

Concurrently, machine learning (ML) has evolved exponentially in solving many complex real-world tasks because it can learn underlying patterns of related components from associated data~\cite{Aurelien:19}. However, its application in strong-field physics and ultrafast science is still in its early stages. Recently, there has been a growing number of studies where ML is used to reduce intensive \textit{ab initio} methods for calculating nonlinear spectra~\cite{Liu:prl20,Yan:oe22,Lytova:CanJP23,Pablos:cpc23}, inversely recovering pulse or molecular parameters~\cite{Zahavy:Optica18,Liu:ComChem21,Klimkin:oe23,Shvetsov:pra23}, and tracking molecular dynamics~\cite{He:pra19,He:NatCom22}. ML offers an alternative approach for solving the Schr\"odinger equation for given potential models or molecular geometries~\cite{Mills:pra17,Hermann:NatChem20,Frank:AnnRe:20}. However, to our knowledge, its applications for the inverse problems, i.e., recovering potentials from observables obtained from the Schr{\"o}dinger equation, such as orbital energies, symmetries, and electric dipole moments, are yet to be explored. These physical data for atomic or molecular orbitals can be achieved from experimental measurements or \textit{ab initio} simulations. Characterizing molecular potentials is crucial for simulating molecular features or spectra emitted from the interaction between molecules and external fields, particularly during molecular dynamics when the molecular configuration changes over time. 

The goal of this paper is twofold. \underline{The first} is to develop an ML engine to construct the molecular potential from a given set of molecular features, including electron energies, dipole moments, and symmetries. The created molecular potential is required to reproduce precise energies and other molecular features for not only HOMO but also the lower-lying orbitals of polar molecules such as HOMO-1 and HOMO-2. \underline{The second} is to apply the achieved molecular potential to a molecular dynamics process induced by the laser-molecule interaction where the role of lower-lying orbitals can not be neglected. 

Since the electronic molecular potential is highly sensitive to the potential parameters, any minor adjustment would cause a noticeable difference in the spatial gradient near the Coulomb singularities, significantly impacting molecular features. To construct the molecular potential from molecular features, we leverage the advantage of ML in computer vision tasks by using high-performance models, such as Convolutional Neural Network (CNN)~\cite{O'Shea:arXiv15} and DeConvolutional Neural Network (DCNN)~\cite{Yang:arXiv21}. Hence, we design the first goal in two steps: (i) To build ML models using DCNN architecture to predict the image of soft-Coulomb-type pseudo-potential within the single-active electron approximation from the set of molecular features (We refer to those as SAE ML models); (ii) To build a ML model named pic2para using the CNN architecture to convert the potential image into analytical form with appropriate potential parameters. 

To achieve the given objectives, we prepare data sets with more than a hundred thousand data points by numerically solving the time-independent Schr\"odinger equation (TISE) with the HCN molecular potentials of various molecular configurations. These generated data sets are then used to train the SAE-POT-ML model. After training, the predicted potentials are used to simulate HHG during the interaction between an HCN molecule and an intense linearly polarized laser pulse by numerically solving the time-dependent Schr\"{o}dinger equation (TDSE). Since the wave functions are mostly spread in the laser polarization direction, we employed the two-dimensional (2D) model. The dimensionality reduction is widely applied to simulate nonlinear phenomena in strong-field physics to save computational costs while preserving their main features~\cite{Lein:pra02,Chirilac:pra10,Shvetsov:jpb21,Shvetsov:pra22,Shvetsov:pra23,Zhang:pra23,Long:jpb23}. 

It is noteworthy that for probing molecular dynamics, it is necessary to repeatedly construct molecular potentials for each molecular arrangement. Thus, we show the substantial outperformance of the transfer learning technique in building the ML potentials for each molecular configuration during its dynamic process. The potential model for a hydrogen cyanide (HCN) molecule is chosen as a paradigmatic example since fingerprints of molecular lower-lying orbitals in HHG have been reported for many molecules but rarely for HCN~\cite{Chu:pra23, Koushki:jmm23}, which has more room for further consideration, especially its dynamic processes. 

The success in constructing the ML potentials inspires us to move forward to applying them to compute and then investigate the signature of lower-lying orbitals in the HHG emitted from an HCN molecule during the H-CN stretching process, i.e., the movement of H atom (the lightest atom of HCN) along the molecular axis. We confirm the imprint of HOMO-1 in the spectral HHG of the HCN molecule, as similarly discovered for N$_2$~\cite{McFarland:Sci08} and CO$_2$~\cite{Smirnova:Nat09}. This result is consistent with other works for HCN~\cite{Chu:pra23,Koushki:jmm23}. Also, we discover the effect of HOMO-1 in the time-frequency profile of the HHG. Furthermore, we uncover the changing of the plateau structure and smoothness of the HHG during the movement of the H atom.        

The rest of the paper is organized as follows. In Sec.~\ref{Sec2}, we briefly describe the concept of the SAE potential model for HCN molecules. In Sec.~\ref{Sec3}, we present the ML procedure to construct the SC SAE potentials from molecular features of HCN during the movement of the hydrogen atom in the H-CN stretching process. The applications of constructed ML SAE potentials in diagnosing the imprints of HOMO-1 in HHG from HCN molecule during the movement of the hydrogen atom are presented in Sec.~\ref{Sec4}. The paper is finished with the conclusions. Atomic units are utilized throughout the paper unless stated otherwise.

%%=====================================================================================================================================================
%% SECTION 2: Single-active-electron potential for HCN molecule
%%=====================================================================================================================================================

\section{Single-active-electron potential for HCN molecule}
\label{Sec2}

In the framework of the SAE model, a molecule can be described by a system of one electron moving in an effective potential called the SAE potential. In this case, the Schr\"{o}dinger equation can be written as
%%--------------------------------------------------------------------------------------------------%%
%% EQ 1: TISE
%%--------------------------------------------------------------------------------------------------%%
\begin{equation}
\left\{-\frac{1}{2}\left(\frac{\partial^2}{\partial x^2}+\frac{\partial^2}{\partial y^2}\right)+V_{\text{SAE}}(x,y)
-E \right\}\psi(x, y)=0,
\label{eq:TISE}
\end{equation}
%%--------------------------------------------------------------------------------------------------%%
%%--------------------------------------------------------------------------------------------------%%
where the 2D model is employed with the active electron position ${{\mathbf{r}}=(x,y)}$ and the SAE potential $V_{\text{SAE}}(x, y)$. $E$ is an electron energy.

There are several ways to construct the SAE potential~\cite{Reiff:JoPC20,Peters:pra12, Nalda:pra04, Abu-samha:pra10,Abu:pra20,Zhao:pra10,Chen:prl13,Sun:pra20,Long:jpb23, Zhu:pra15,Wang:pra20}; however, in the present study, we are interested in the soft-Coulomb-type pseudo-potential~\cite{Nalda:pra04,Peters:pra12,Chen:prl13,Sun:pra20,Zhu:pra15,Long:jpb23,Wang:pra17, Wang:pra20}, which has been successfully applied to diatomic or symmetric triatomic molecules such as CO$_2$~\cite{Peters:pra12,Nalda:pra04,Zhu:pra15}, CO~\cite{Chen:prl13,Wang:pra17,Sun:pra20}, N$_2$~\cite{Zhu:pra15,Peters:pra12,Long:jpb23}, HeH$^{+}$~\cite{Wang:pra20}, and BF~\cite{Wang:pra17}. More recently, it has also been applied to investigate the multiple-orbital contribution in HHG emitted from an N$_2$ molecule~\cite{Long:jpb23}. We generalize the potential model for our interested molecule, asymmetric triatomic molecule HCN, as
%%--------------------------------------------------------------------------------------------------%%
%% EQ 2: V_SAE
%%--------------------------------------------------------------------------------------------------%%
\begin{equation}
V_{{\text{SAE}}}(\mathbf{r})=\sum_{\alpha=\mathrm{H,C,N}}\frac{-Z_\alpha(\mathbf{r})}{\sqrt{|\mathbf{r}-\mathbf{R}_\alpha|^2+a_\alpha^2}} .
\label{eq:V_SAE}
\end{equation}
%%--------------------------------------------------------------------------------------------------%%
%%--------------------------------------------------------------------------------------------------%%
Here, $\alpha$ is the label of each molecular nuclei (H, C, N) located at positions $\mathbf{R}_{\alpha}$; parameters $a_\alpha$ are used to soften the Coulomb interaction. The position-dependent screened effective charges are defined by
%%--------------------------------------------------------------------------------------------------%%
%% EQ 3: effective-charge
%%--------------------------------------------------------------------------------------------------%%
\begin{equation}
Z_\alpha(\mathbf{r})=Z_\alpha^\infty+(Z_\alpha^0-Z_\alpha^\infty)\,\text{exp}\left[-\frac{|{\bf{r}}-{\bf{R}}_\alpha|^2} {{\sigma_\alpha}^2}\right],
\label{eq:effective-charge}
\end{equation}
%%--------------------------------------------------------------------------------------------------%%
%%--------------------------------------------------------------------------------------------------%%
where parameters $\sigma_\alpha$ are introduced to account for the distance-dependent electron-electron screening effect. $Z_{\alpha}^0$ is the bare charge of nucleus $\alpha$, while $Z_\alpha^\infty$ denotes its effective charge as seen by the electron at infinity. The bare charges of nuclei are known as $Z_\mathrm{H}^0=+ 1$~a.u., $Z_\mathrm{C}^0=+ 6$~a.u., and $Z_\mathrm{N}^0=+ 7$~a.u. The effective charges are $Z_\mathrm{H}^\infty=+ 1.0/14.0$~a.u., $Z_\mathrm{C}^\infty= + 6.0/14.0$~a.u., and $Z_\mathrm{N}^\infty= + 7.0/14.0$~a.u. We note that some authors derived the effective charges from the Mulliken analysis \cite{Peters:pra12}; however, we prefer the simple model following the work of Chen \textit{et al.}~\cite{Chen:prl13}.

%%--------------------------------------------------------------------------------------------------%%
%% FIG 1: HCN
%%--------------------------------------------------------------------------------------------------%%
\begin{figure}[htb]
	\begin{center}
		\includegraphics[width=1.0\linewidth]{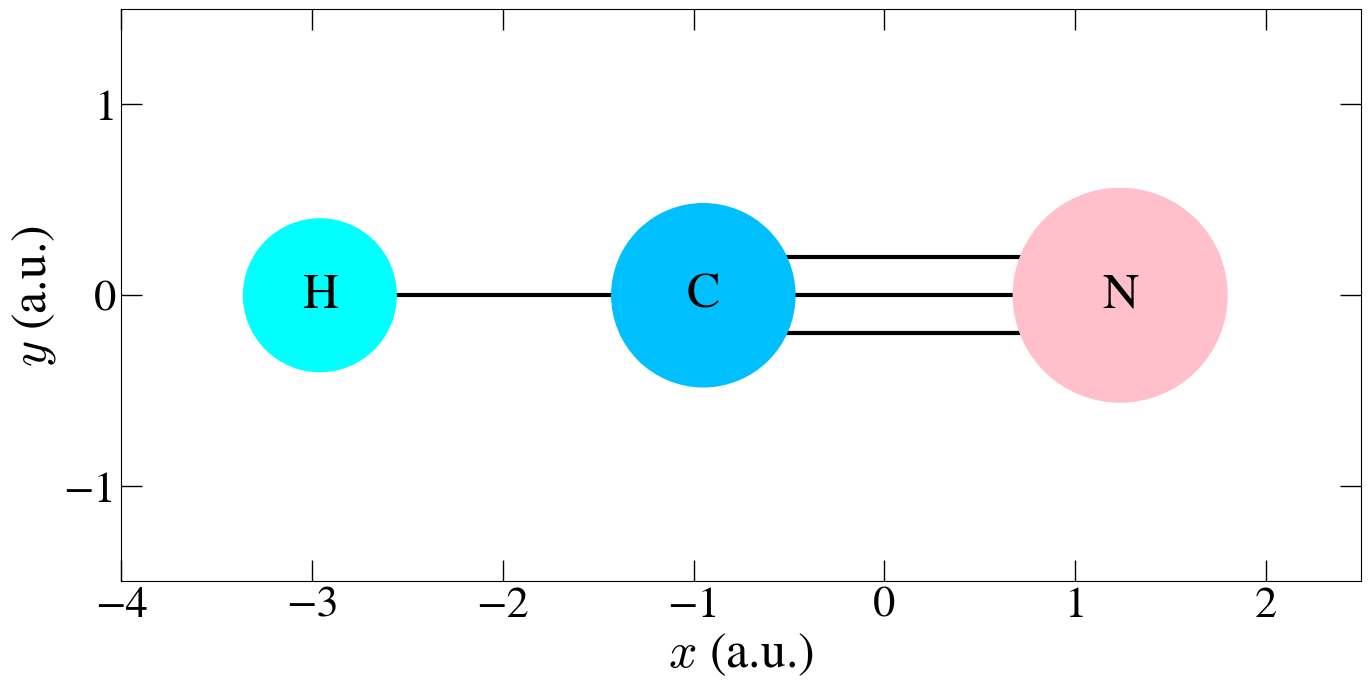}
	\end{center}
	\caption{Sketch of 2D HCN molecule at the equilibrium. Here, the molecular charge center is placed at the origin. The positions of atoms H, C, and N are respectively $x_\mathrm{H}=-2.959$~a.u., $x_\mathrm{C}=-0.949$~a.u., and $x_\mathrm{N}=+1.236$~a.u. The interatomic separations H-C $=2.010$~a.u. and C-N $=2.185$~a.u. are calculated by the GAUSSIAN~09 code.}
\label{fig:HCN}
\end{figure}
%%--------------------------------------------------------------------------------------------------%%
%%--------------------------------------------------------------------------------------------------%%

Figure~\ref{fig:HCN} shows a sketch of the 2D HCN molecule at equilibrium, with the molecular charge center fixed at the origin of the coordinates and the molecular axis meticulously aligned on the $x$-axis. The positions of H, C, and N nuclei at equilibrium are $x_\mathrm{H}=-2.959$~a.u., $x_\mathrm{C}=-0.949$~a.u., and $x_\mathrm{N}=+1.236$~a.u., respectively. These values, obtained from the optimization job using the Hartree-Fock method implemented in the GAUSSIAN~09 software~\cite{g09}, are consistent with the experimental data and other theoretical calculations~\cite{Tanja:chemphys01, koput:cpl96}.
In considering the H-CN stretching process, coordinate $x_\mathrm{H}$ varies from $-2.4$~a.u. to $-4.0$~a.u., according to the turning points when nuclei are excited to the total energy level by 2.8~eV higher than the energy minimum of the \mbox{C-H} stretch potential curve. Due to the much heavier nuclear masses of N and C, we assume that their positions are fixed and adiabatically unaffected by the variation of the H atom.  

Six parameters that remain undefined in the model potential~\eqref{eq:V_SAE} are $a_\mathrm{H}$, $a_\mathrm{C}$, $a_\mathrm{N}$, $\sigma_\mathrm{H}$, $\sigma_\mathrm{C}$, and $\sigma_\mathrm{N}$. They are chosen so that the SAE potential can properly describe the HCN molecule, meaning that its energies, dipole moments, and symmetries for HOMO, HOMO-1, and HOMO-2 are comparable with those calculated by the {\it{ab initio}} method. We note that in some previous studies for HHG calculations, where it is interested only in the HOMO~\cite{Xu:pra09,Peters:pra12}, the sets of optimized parameters are not unique. Therefore, people freely pick up an appropriate set of parameters to get the right HOMO energy of the molecule. However, some recent tasks require not only an adequate HOMO but also accurate HOMO-1, HOMO-2, and even deeper-lying orbitals since the possibility of orbital coupling in nonlinear physical phenomena~\cite{Shu:prl22,Camper:prl23,Fu:oe23,Luppi:JCPA23,Chu:pra23,Long:jpb23}. Moreover, for a polar molecule such as HCN, the interested physical quantities are restricted not only by energies and orbital symmetries but also by the orbital asymmetries characterized by electric dipole moments. Furthermore, for polyatomic molecules consisting of various kinds of atoms, searching for potential parameters should follow the spirit of the theory of linear combination of atomic orbitals (LCAO), i.e., molecular orbitals are formed from reasonable combinations of atomic orbitals from constitute atoms. 

To satisfy the rigorous necessities mentioned above, constructing the model potential by strictly choosing six parameters requires a huge amount of data to analyze. In this case, scanning data to get a set of six desired potential parameters is impractical. Other conventional optimization methods, such as genetic algorithms, are challenging to implement due to their high computational demands and time consumption. In the next section, we build a model based on machine learning to construct the model potential for molecule HCN. 

%%=====================================================================================================================================================
%% SECTION 3: Machine-learning-based SAE potential model for HCN molecule
%%=====================================================================================================================================================

\section{Machine-learning-based SAE potential model for HCN molecule}
\label{Sec3}

In this Section, we build a machine-learning model for the soft-Coulomb-type SAE potential of an HCN molecule, which predicts six potential parameters from input data consisting of energies, electric dipoles, and symmetries of molecular orbitals. We call it the SAE-POT-ML model. The input data can be taken from experiments or simulations by chemical codes such as GAUSSIAN or GAMESS. The model is constructed to work for the polar triatomic molecule HCN with fixed positions of N and C, but the H position is in equilibrium or varies in the range in Table~\ref{tab:parameters}. In the following, we describe the SAE-POT-ML model's construction and general working scheme, which consists of three main phases, as presented in Fig.~\ref{fig:main_pro}. The details are enclosed in the Supplementary Material.

%%--------------------------------------------------------------------------------------------------%%
%% FIG 2: main_pro
%%--------------------------------------------------------------------------------------------------%%
\begin{figure*}[htb]
	\begin{center}
		\includegraphics[width=0.85\linewidth]{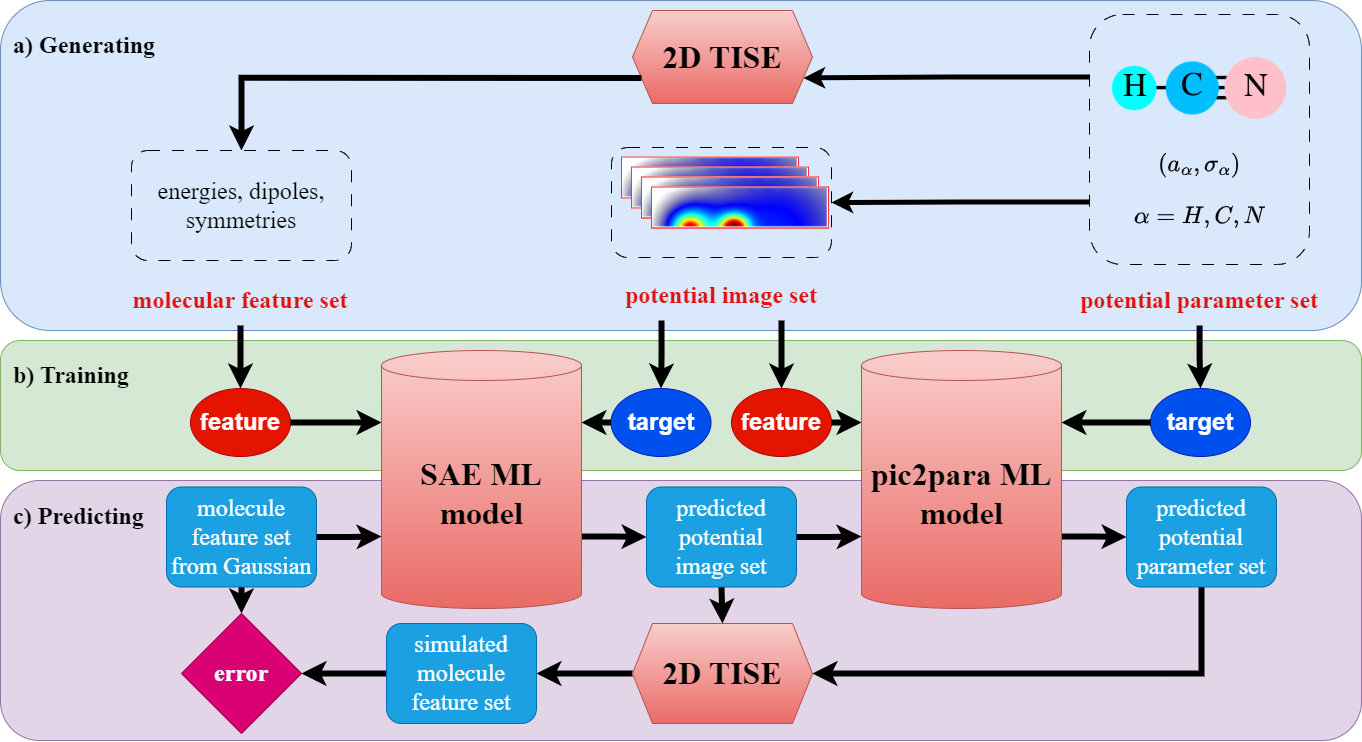}
	\end{center}
	\caption{Construction and general working scheme of the SAE-POT-ML model, consisting of three phases: (a)~generating datasets, (b)~training ML models, and (c)~application in predicting SAE potential. In the first phase (a), the data set comprising the 2D potential (image or parameters) and the molecular feature set (energies, orbital symmetries, and electric dipole moments for HOMO, HOMO-1, and HOMO-2) are generated by the TISE source code. The 2D potential image is an image converted from the potential parameters. These datasets are served for the training phase (b), where the SAE-POT-ML model is built in two steps: the SAE ML model predicts a 2D potential image (as an output) from the molecular feature set (as an input), then from this 2D potential image the pic2para ML model can extract the potential parameters (as an alternative output). The trained ML models are then applied to predict SAE potential (c). The accuracy of the predictions is evaluated by benchmarking the simulated molecular feature set, calculated by the TISE code with the predicted SAE potential, and that of input simulated from GAUSSIAN~09.\\}
\label{fig:main_pro}
\end{figure*} 

%%--------------------------------------------------------------------------------------------------%%
%% TAB 1: parameters
%%--------------------------------------------------------------------------------------------------%%
\begin{table}[htb] 
    \centering
    \caption{Ranges of the potential parameters (in atomic units) and nuclear positions for the HCN molecule considered in this study.}
    
    \begin{tabular}{
     >{\centering\arraybackslash}m{0.5cm}
    |>{\centering\arraybackslash}m{2.0cm}
    |>{\centering\arraybackslash}m{2.0cm}
    |>{\centering\arraybackslash}m{2.0cm}
    }                                          
                    & H                     & C                     & N                     \\ \Xhline{1 pt}
    $a_\alpha$      & [0.1,~0.5]            & [0.8,~3.2]            & [3.0,~10.0]           \\ 
    $\sigma_\alpha$ & [1.0,~3.0]            & [0.23,~2.41]          & [1.04,~8.60]          \\ \hline
    $x_{\alpha}$    & [$-4.0$,~$-2.4$]      & $-0.949$              &  $+1.236$             \\ 
    \end{tabular}
    
\label{tab:parameters}
\end{table}
%%--------------------------------------------------------------------------------------------------%%
%%--------------------------------------------------------------------------------------------------%%

%%--------------------------------------------------------------------------------------------------%%
%% Generating datasets
%%--------------------------------------------------------------------------------------------------%%

{\bf{Generating datasets} --} Each data point consists of a potential set (six parameters of the potential) and a molecular feature set (orbital energies, electric dipole moments, and orbital symmetries). We prepare the dataset for training the ML model by numerically solving the time-independent Schr\"{o}dinger equation with various potential sets of $a_\alpha$ and $\sigma_\alpha$, whose ranges are shown in Table~\ref{tab:parameters}. Among many effective methods, we choose the imaginary time propagation method~\cite{Kosloff:cpc86} and prepare a source code that gives convergent energies, electric dipole moments, and orbital symmetries to generate the datasets. Here, we are interested in only three orbitals: HOMO, \mbox{HOMO-1}, and \mbox{HOMO-2}. 

ML models can learn better if the training data contains enough relevant features. Besides, with the same amount of data, the narrow range of parameters aids the ML model in learning more effectively than the wide range. Therefore, feature engineering is essential to get a high-quality dataset in training an ML model. To this end, we found the appropriate ranges of potential parameters for generating the training sets, as shown in Table~\ref{tab:parameters}. Here, the ranges of $\sigma_{\mathrm{C}}$ and $\sigma_{\mathrm{N}}$ are restricted by the power or linear expressions of $a_{\mathrm{C}}$ and $a_{\mathrm{N}}$ correspondingly (not shown). These ranges are roughly chosen based on the spirit of LCAO, i.e., based on the initial estimation of possibilities for forming molecular orbitals from appropriate atomic orbitals whose energies are calculated by solving the TISE of the bare atoms. As a result, we obtained a dataset with 30,000 data points for the HCN molecule with H at the equilibrium position and with 100,000 for H at various positions and used it to train the SAE-POT-ML model. This number of data points for a wide range of H atom's positions is considerably small compared to that of one fixed position at equilibrium thanks to the transfer learning based on the pre-trained ML model of the HCN with fixed H position.  

%%--------------------------------------------------------------------------------------------------%%
%% SAE-POT-ML model's construction
%%--------------------------------------------------------------------------------------------------%%

{\bf {SAE-POT-ML model's construction} --} We initially attempted to develop an ML potential model that can directly and independently predict the values of $\sigma_\alpha$ and $a_\alpha$ from the molecular feature set. However, the model's accuracy was compromised because of the need for a correlation between each value in the potential set. In addition, due to the high sensitivity, any minor adjustment of potential parameters would cause a noticeable difference in the spatial gradient near the Coulomb singularities, significantly impacting molecular features. In response to this shortcoming, we take advantage of ML in imaging recognition and, thus, develop a two-step ML potential model. (i)~In the first step, we convert the potential parameters to the 2D potential image, correlating all values of the potential parameter set together. Then, we develop the SAE ML model to predict the 2D potential image (instead of the potential parameter set) from the molecular feature set. (ii)~In the second step, the pic2para ML model is built to extract the potential parameters ($\sigma_\alpha$ and $a_\alpha$) from the predicted 2D potential image. 

For the SAE ML model, we use the DCNN to build a model based on the molecular feature set (energies, dipole moments, and orbital symmetries) to predict the 2D potential images. To improve the effectiveness of the SAE ML model with a moderate dataset, we do not directly train the SAE ML for both two cases of fixed and moving H atoms but in two stages. First, the preSAE ML model is trained to construct the 2D potential images from the molecular feature set of the HCN molecule at equilibrium. Then, to reduce the computational cost, transfer learning is applied to retrain the preSAE ML model with the new dataset of the HCN molecule (with moving H atom) to predict the potential images in the \mbox{H-CN} stretching process. For the pic2para model, it should be noted that the 2D potential image is easy to construct with the potential parameter set using Eq.~\eqref{eq:V_SAE}. However, the reversed action is non-trivial. So, we apply the CNN to build the pic2para ML model that extracts the set of potential parameters from the 2D potential images. For all models, we employed multiple techniques (cross-validation~\cite{wolpert1992stacked,sollich1995learning}, early stopping~\cite{prechelt2002early}, dropout layer~\cite{srivastava2014dropout}) during the model-building process to prevent overfitting. Indeed, the loss function drops smoothly, and its values are acceptably small, as shown in Supplementary Material, which confirms our model is well constructed.

\begin{figure}[htb!]
	\begin{center}
		\includegraphics[width=1.0\linewidth]{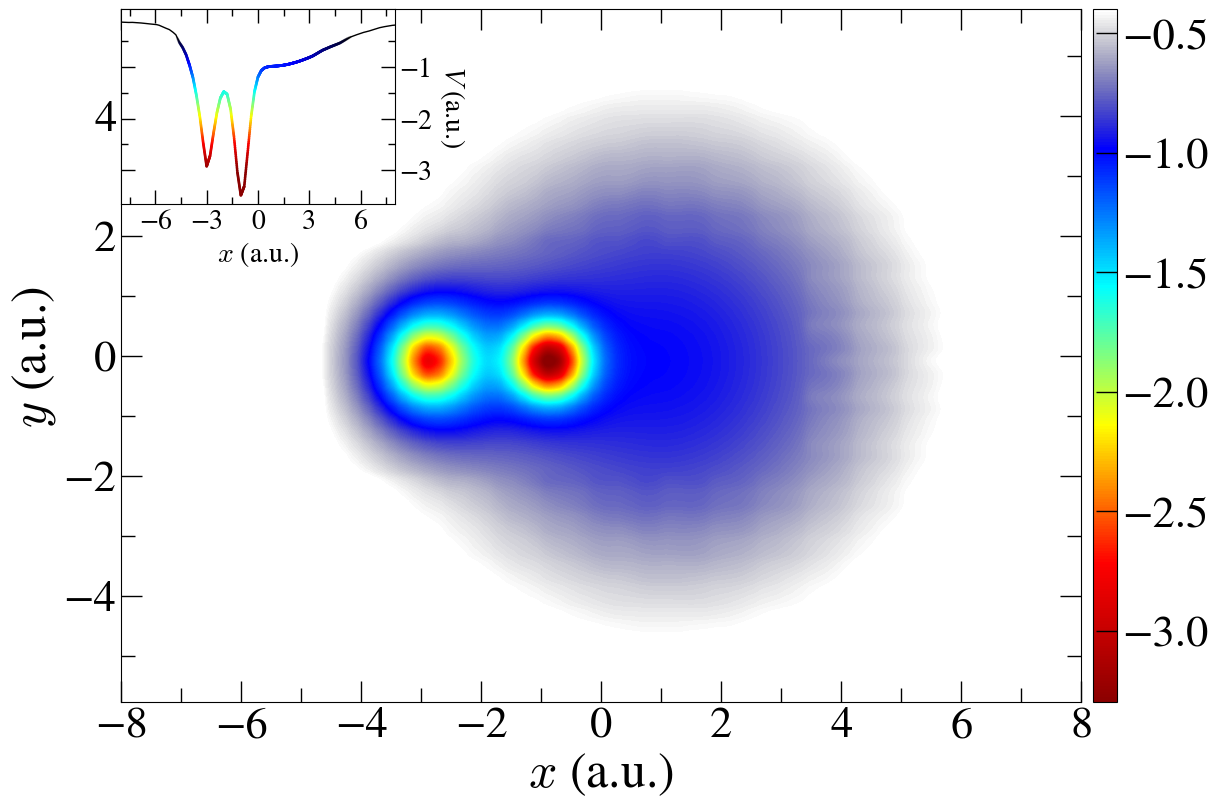}
	\end{center}
	\caption{The 2D SAE potential image of the molecule HCN at the equilibrium predicted by the SAE-POT-ML model from the molecular feature set yielded by the GAUSSIAN code. Here, we present only the most information-rich part of the potential, but the full image with the $x$-axis from -20 to +20~a.u. and $y$-axis from -19.8 to +19.8~a.u is used in the training model and prediction for better accuracy. The inset figure presents a slice at $y = 0$.}
\label{fig:V_SAE}
\end{figure}

{\bf {Prediction of SAE potential for HCN molecule} --} From the built SAE-POT-ML model, we can theoretically predict the SAE potential of the HCN molecule in the form of Eq.~\eqref{eq:V_SAE} from the molecular feature sets (experimental or simulated theoretically). In this study, we use the GAUSSIAN~09 code with the Hartree-Fock method and the 6-311+g(2d,p) basis set to prepare molecular feature sets for the HCN molecule and call them the \textit{true sets} for brevity. From these data, our ML model predicts the 2D potential image, as shown in Fig.~\ref{fig:V_SAE}, for an example of the HCN molecule at equilibrium. From these 2D potential images, the pic2para ML model extracts the analytical form \eqref{eq:V_SAE} of the SAE potential, whose potential parameters for the equilibrium HCN are also shown in Table~\ref{tab:pred_para_E}. 

%%--------------------------------------------------------------------------------------------------%%
%% Prediction of SAE potential for HCN molecule
%%--------------------------------------------------------------------------------------------------%%

%%--------------------------------------------------------------------------------------------------%%
%% FIG 3: V_SAE
%%--------------------------------------------------------------------------------------------------%%
%%--------------------------------------------------------------------------------------------------%%
%%--------------------------------------------------------------------------------------------------%%

%%--------------------------------------------------------------------------------------------------%%
%% TAB 2: pred_para_E
%%--------------------------------------------------------------------------------------------------%%
\begin{table}[h!] %\fontsize{8.5}{9.5}\selectfont 
    \centering 
    \caption{The SAE potential parameters in atomic units predicted by the SAE-POT-ML model from the molecule feature sets yielded by the GAUSSIAN code for an HCN molecule at equilibrium.} 
    
    \begin{tabular}{
     >{\centering\arraybackslash}m{1.11cm}
    |>{\centering\arraybackslash}m{1.11cm}
     >{\centering\arraybackslash}m{1.11cm}
    |>{\centering\arraybackslash}m{1.11cm}
     >{\centering\arraybackslash}m{1.11cm}
    |>{\centering\arraybackslash}m{1.11cm}
     >{\centering\arraybackslash}m{1.11cm}
    }  

    $x_\mathrm{H}$ &
    $a_\mathrm{H}$ & $\sigma_\mathrm{H}$ &
    $a_\mathrm{C}$ & $\sigma_\mathrm{C}$ &
    $a_\mathrm{N}$ & $\sigma_\mathrm{N}$ \\
    \Xhline{1pt}
    
    -2.959 &
    0.3696 & 1.1600 &
    2.4455 & 0.8050 &
    8.6025 & 4.7652 \\
    
    \end{tabular}
    
\label{tab:pred_para_E}
\end{table}
%%--------------------------------------------------------------------------------------------------%%
%%--------------------------------------------------------------------------------------------------%%

%%--------------------------------------------------------------------------------------------------%%
%% TAB 3: equil
%%--------------------------------------------------------------------------------------------------%%
\begin{table}[t] %\fontsize{9.5}{10}\selectfont
    \centering
    \caption{The input (from GAUSSIAN~09) (true) and output (calculated by the TISE code using the SAE potential parameters predicted from the SAE-POT-ML model) (pred.) molecular feature sets of the HCN molecule in equilibrium.}
    
    \begin{tabular}{
     >{\centering\arraybackslash}m{2.5cm}
    |>{\centering\arraybackslash}m{1.8cm}
    |>{\centering\arraybackslash}m{1.8cm}
    |>{\centering\arraybackslash}m{1.8cm}
    }      
     & HOMO & HOMO-1 & HOMO-2               \\ 
    \Xhline{1pt}
    
    \makecell[b]{Spatial \\ distribution \vspace{0.35cm}} &
    \includegraphics[width=0.8\linewidth]{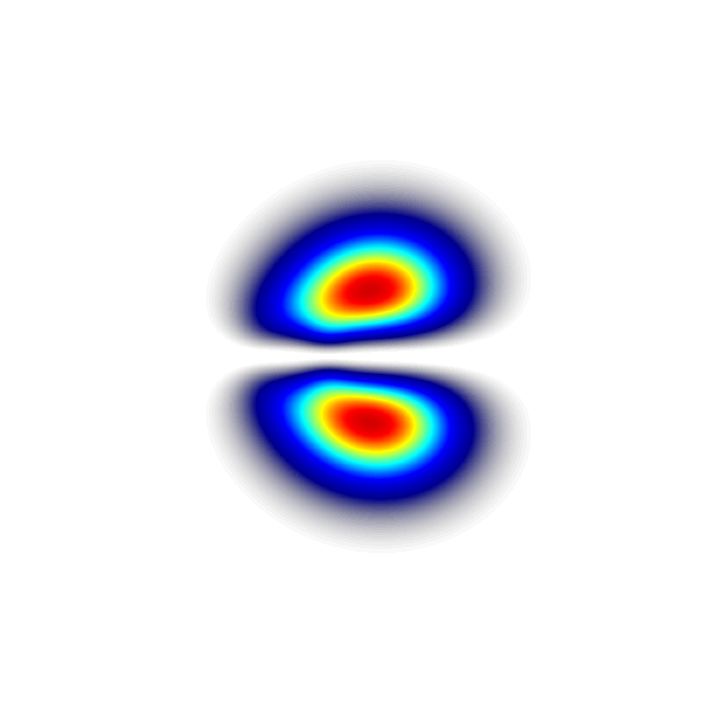} &
    \includegraphics[width=0.8\linewidth]{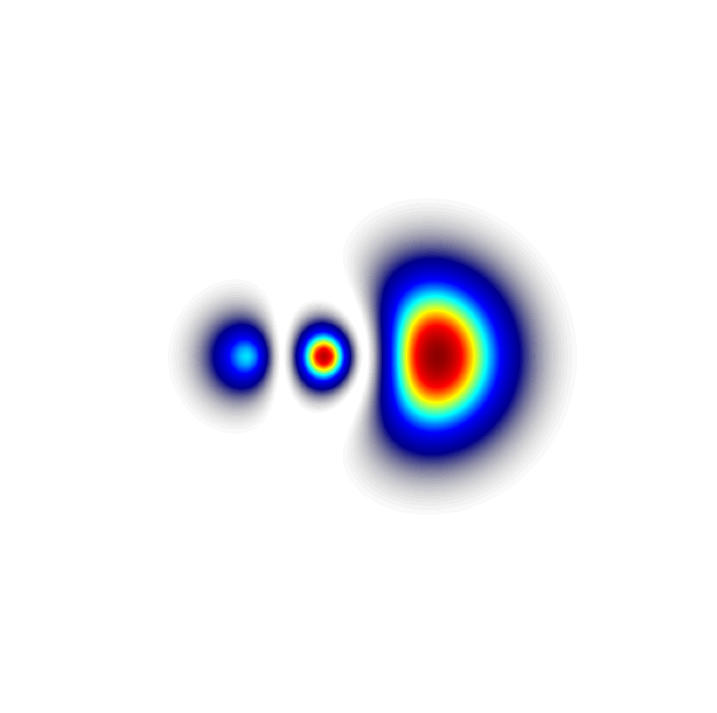} &
    \includegraphics[width=0.8\linewidth]{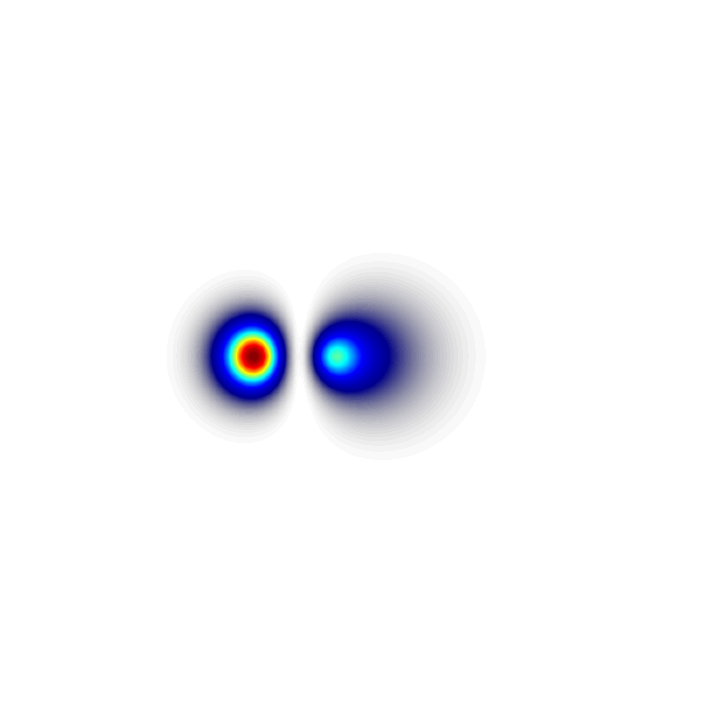}                          \\
    \hline
    
    $E$ (true)  &                -0.4964	&       -0.5839 &       -0.8159     \\
    $E$ (pred.) &                -0.4985    &       -0.5849	&       -0.8310	    \\
    $|\Delta E|/E$~(true) &       0.42~\%   &        0.16~\%&        1.85~\%    \\
    \hline
    
    $d$ (true)  &                -0.2421	&       -1.5161	&        1.9199     \\
    $d$ (pred.) &                -0.2439	&       -1.5829	&        1.9091     \\
    $|\Delta d|/d$~(true) &       0.73~\%   &        4.41~\%&        0.56~\%    \\
       
    \end{tabular}
    
\label{tab:equil}
\end{table}
% \end{table}
%%--------------------------------------------------------------------------------------------------%%
%%--------------------------------------------------------------------------------------------------%%

To justify the accuracy of the predicted SAE potential for HCN at equilibrium, we calculate the molecular features by the TISE code with this potential, either in image or analytical forms. We call the set of these calculated features the \textit{predicted set} or the output and then compare it with the \textit{true set}, the input molecular feature set from the GAUSSIAN~09 code. The orbital symmetries for the input and output coincide with each other. Also, the results for energies and dipole moments are displayed in Table~\ref{tab:equil}. The comparison in this table demonstrates the highly accurate performance of the SAE-POT-ML model in predicting orbital symmetries, energies, and dipole moments of the HOMO, \mbox{HOMO-1}, and \mbox{HOMO-2}. Indeed, the differences between the \textit{predicted} and the \textit{true} values are less than 1.85~\% for energy and 4.41~\% for dipole. This accuracy is comparable and even higher than that of other SAE potentials, which are popularly applied in strong-field physics~\cite{Zhao:pra10,Abu-samha:pra10,Peters:pra12,Abu:pra20}. We note that the relative errors of energies are in the same order as those when using potential in the image form. At the same time, dipole moments' errors slightly increase compared to those obtained by the potential in the analytical form. In short, the constructed SAE-POT-ML model can predict an accurate SAE potential in both image and analytical forms that correctly captures the asymptotic behavior of the potential from the \textit{ab initio} calculations for the HCN molecule in the equilibrium position of H. 

%%--------------------------------------------------------------------------------------------------%%
%% TAB 4: pred_para1
%%--------------------------------------------------------------------------------------------------%%
\begin{table}[h!] % \fontsize{8.5}{9.5}\selectfont
    \centering 
    \caption{SAE potential parameters in atomic units predicted by the SAE-POT-ML model from the molecular feature sets yielded by the chemical code GAUSSIAN~09 for the HCN molecule with various H positions stretched from the equilibrium, i.e., $x_\mathrm{H}<-2.96$ a.u. (\mbox{H-C~$>2.01$~a.u.)}.}
  
    \begin{tabular}{
     >{\centering\arraybackslash}m{1.11cm}
    |>{\centering\arraybackslash}m{1.11cm}
     >{\centering\arraybackslash}m{1.11cm}
    |>{\centering\arraybackslash}m{1.11cm}
     >{\centering\arraybackslash}m{1.11cm}
    |>{\centering\arraybackslash}m{1.11cm}
     >{\centering\arraybackslash}m{1.11cm}
    }  

    $x_\mathrm{H}$ &
    $a_\mathrm{H}$ & $\sigma_\mathrm{H}$ &
    $a_\mathrm{C}$ & $\sigma_\mathrm{C}$ &
    $a_\mathrm{N}$ & $\sigma_\mathrm{N}$ \\ \Xhline{1pt}
    
    -3.88 & 0.4646 & 1.1594 & 3.0648 & 1.0296 & 8.5690 & 4.3546 \\
    -3.84 & 0.4663 & 1.1628 & 3.0468 & 1.0173 & 8.5856 & 4.3932 \\
    -3.80 & 0.4663 & 1.1400 & 3.0245 & 1.0045 & 8.5956 & 4.4311 \\
    -3.76 & 0.4669 & 1.1300 & 3.0012 & 0.9909 & 8.6065 & 4.4733 \\
    -3.72 & 0.4685 & 1.1600 & 2.9714 & 0.9754 & 8.6087 & 4.5260 \\
    -3.68 & 0.4720 & 1.1659 & 2.9350 & 0.9599 & 8.6040 & 4.5630 \\
    -3.60 & 0.4691 & 1.1300 & 2.8104 & 0.9363 & 8.6409 & 4.6560 \\
    -3.56 & 0.4719 & 1.1100 & 2.7385 & 0.9232 & 8.6819 & 4.7012 \\
    -3.52 & 0.4743 & 1.0800 & 2.6786 & 0.9120 & 8.7202 & 4.7557 \\
    -3.48 & 0.4756 & 1.0600 & 2.6443 & 0.9019 & 8.7592 & 4.8132 \\
    -3.44 & 0.4770 & 1.0600 & 2.6237 & 0.8914 & 8.7864 & 4.8675 \\
    -3.40 & 0.4782 & 1.1300 & 2.6045 & 0.8778 & 8.8093 & 4.8830 \\
    -3.36 & 0.4780 & 1.1693 & 2.5605 & 0.8595 & 8.8240 & 4.8901 \\
    -3.32 & 0.4774 & 1.1861 & 2.5111 & 0.8409 & 8.8321 & 4.9223 \\
    -3.28 & 0.4789 & 1.2500 & 2.4900 & 0.8228 & 8.8318 & 4.9774 \\
    -3.24 & 0.4800 & 1.2900 & 2.4775 & 0.8031 & 8.8247 & 5.0382 \\
    -3.20 & 0.4796 & 1.3500 & 2.4738 & 0.7818 & 8.8213 & 5.0713 \\
    -3.16 & 0.4772 & 1.4000 & 2.4883 & 0.7569 & 8.8120 & 5.0697 \\
    -3.12 & 0.4760 & 1.4800 & 2.5411 & 0.7297 & 8.7612 & 5.0391 \\
    -3.08 & 0.4677 & 1.5500 & 2.5699 & 0.7026 & 8.6968 & 4.9942 \\
    -3.04 & 0.4516 & 1.6000 & 2.5612 & 0.6749 & 8.6394 & 4.9542 \\
    -3.00 & 0.4322 & 1.6100 & 2.5257 & 0.6493 & 8.5850 & 4.8608 \\ 
    
    \end{tabular}
    
\label{tab:pred_para1}
\end{table}
%%--------------------------------------------------------------------------------------------------%%
%%--------------------------------------------------------------------------------------------------%%

It should be noted that after 100 hours of generating data and 6 hours for training, the SAE-ML-POT model can predict potential parameters within three dozen centiseconds, which shows superior performance compared to the traditional optimized algorithm. Indeed, we also tried the prediction with the genetic algorithm with identical settings (with the generating phase for the SAE-ML-POT model), which resulted in the best match taking 60 hours to obtain parameters for just one position. However, the precision was significantly lower compared to the SAE-ML-POT model.

Once the transfer learning method is implemented, the constructed SAE-POT-ML model can also predict the SAE potential for the molecule HCN when the H nucleus is displaced from the equilibrium position on the $x$-axis. Indeed, we provide the SAE potential parameters for numerous cases in Tables \ref{tab:pred_para1} and \ref{tab:pred_para2}. It should be noted that after preliminary examining the HHG emitted from HCN at equilibrium (see Section~\ref{Sec4} for details), we found that only HOMO and HOMO-1 contributed to HHG, while HOMO-2 and lower-lying are not evolved. Therefore, to improve the performance of the SAE-POT-ML model and reduce the computational cost of a large training dataset, we eliminate the requirement of the dipole moment of HOMO-2 in the molecular feature set in the training phase. 

With the predicted potential parameters shown in Tables \ref{tab:pred_para1} and \ref{tab:pred_para2}, we also justify their accuracy by calculating molecular feature sets via solving the TISE. Since the HOMO's dipole moments are considerably tiny, their values are extremely sensitive to the parameters involved in constructing electron density. In the HCN case, the concerning parameter is $\sigma_\mathrm{H}$, which causes the sensitivity of the dipole moments even to the second digit after decimal. This parameter accounts for the distance-dependent electron-electron screening effect, especially for a large deviation of $x_\mathrm{H}$ from equilibrium. Therefore, after obtaining the predicted potential parameters from the SAE-POT-ML model, we implement minor adjustments of $\sigma_\mathrm{H}$ around its predicted value. These result in much better dipole moments of HOMO but cause a slight trade-off in other features; however, it almost does not change the predicted symmetries and energies.

%%--------------------------------------------------------------------------------------------------%%
%% TAB 5: pred_para2
%%--------------------------------------------------------------------------------------------------%%
\begin{table}[ht!] %\fontsize{8.5}{9.5}\selectfont
    \centering 
    \caption{The same as Table~\ref{tab:pred_para1} but for the HCN molecule with various H positions compressed from the equilibrium, i.e., $x_\mathrm{H}>-2.96$~a.u. (H-C $<2.01$~a.u.).}

    \begin{tabular}{
     >{\centering\arraybackslash}m{1.11cm}
    |>{\centering\arraybackslash}m{1.11cm}
     >{\centering\arraybackslash}m{1.11cm}
    |>{\centering\arraybackslash}m{1.11cm}
     >{\centering\arraybackslash}m{1.11cm}
    |>{\centering\arraybackslash}m{1.11cm}
     >{\centering\arraybackslash}m{1.11cm}
    }  

    $x_\mathrm{H}$ &
    $a_\mathrm{H}$ & $\sigma_\mathrm{H}$ &
    $a_\mathrm{C}$ & $\sigma_\mathrm{C}$ &
    $a_\mathrm{N}$ & $\sigma_\mathrm{N}$ \\ \Xhline{1pt}

    -2.92 & 0.3848 & 1.6300 & 2.5057 & 0.6165 & 8.4716 & 4.7441 \\
    -2.88 & 0.3613 & 1.6300 & 2.4937 & 0.6032 & 8.4335 & 4.7420 \\
    -2.84 & 0.3412 & 1.6200 & 2.4754 & 0.5910 & 8.3744 & 4.7478 \\
    -2.80 & 0.3163 & 1.6200 & 2.4902 & 0.5720 & 8.3001 & 4.7375 \\
    -2.76 & 0.2911 & 1.6500 & 2.5183 & 0.5511 & 8.2203 & 4.6519 \\
    -2.72 & 0.2627 & 1.6500 & 2.5590 & 0.5418 & 8.1642 & 4.6201 \\
    -2.68 & 0.2287 & 1.6300 & 2.6102 & 0.5363 & 8.1189 & 4.5648 \\
    -2.64 & 0.1983 & 1.6300 & 2.7140 & 0.5465 & 8.0471 & 4.4749 \\
    -2.60 & 0.1803 & 1.6100 & 2.8928 & 0.5946 & 7.9856 & 4.3990 \\
    -2.56 & 0.1670 & 1.5300 & 2.9377 & 0.6528 & 7.9514 & 4.3445 \\
    -2.52 & 0.1594 & 1.4200 & 2.9266 & 0.7176 & 7.9026 & 4.2642 \\
    -2.48 & 0.1561 & 1.3100 & 2.8727 & 0.7615 & 7.8534 & 4.1754 \\
    -2.44 & 0.1541 & 1.2100 & 2.7749 & 0.7898 & 7.8545 & 4.0786 \\
    -2.40 & 0.1557 & 1.1500 & 2.7074 & 0.8009 & 7.8316 & 3.9747 \\
    -2.56 & 0.1670 & 1.5300 & 2.9377 & 0.6528 & 7.9514 & 4.3445 \\
    -2.52 & 0.1594 & 1.4200 & 2.9266 & 0.7176 & 7.9026 & 4.2642 \\
    -2.48 & 0.1561 & 1.3100 & 2.8727 & 0.7615 & 7.8534 & 4.1754 \\ 
    -2.44 & 0.1541 & 1.2100 & 2.7749 & 0.7898 & 7.8545 & 4.0786 \\ 

    \end{tabular}
    
\label{tab:pred_para2}
\end{table}
%%--------------------------------------------------------------------------------------------------%%
%%--------------------------------------------------------------------------------------------------%%

For details, solving the TISE with the predicted SAE potential demonstrates exact predictions of orbital symmetries and relatively accurate predicted sets compared to the true sets. The average errors throughout the range of  $x_\mathrm{H}$ are 0.76~\% and 0.51~\% for energies of HOMO and \mbox{HOMO-1}; 4.27~\% and 5.33~\% for dipole moments of HOMO and \mbox{HOMO-1}, respectively. For the energy of HOMO-2, the error values are 1.58~\%. We need to train the ML models with more data for better accuracy. However, the obtained accuracy is sufficient for our purpose, and we apply the dynamic SAE potential of HCN for the application in the next section: The high-order harmonic generation for the HCN molecule with the movement of the H atom. 

Furthermore, since the potential form [Eq.~(\ref{eq:V_SAE})] and construction scheme of the SAE-POT-ML model are general, the SAE-POT-ML model can be generalized to reconstruct soft-Coulomb-type SAE potential for other small molecules with two or three atoms. In these molecules, the constraints and parameters (potential parameter set) are equal to or even less than those of HCN; therefore, the utilization can be straightforward. 
%%=====================================================================================================================================================
%% SECTION 4: Applications: Multiorbital effects in HHG from HCN molecule
%%=====================================================================================================================================================

\section{Applications: Multiorbital effects in HHG from HCN molecule}
\label{Sec4}

This section demonstrates two applications of our predicted SAE potentials from the SAE-POT-ML model in revealing the multiorbital effect in HHGs emitted from the HCN molecule during its dynamic process. 

It is noteworthy that the multiorbital effects in HHG have been uncovered for many atoms~\cite{Worner:prl09,Shiner:NatPhys11} and symmetric molecules~\cite{McFarland:Sci08,Smirnova:Nat09,Li:Sci08,Long:jpb23,Yun:prl15,Shu:prl22,Camper:prl23} but rarely documented for polar molecules. Indeed, only some works have been published recently~\cite{Chu:pra23,Koushki:jmm23,Fu:oe23}, where the influence of deep-lying orbitals on the contrast of odd- and even-orders~\cite{Chu:pra23}, the orientation-dependence~\cite{Koushki:jmm23} in HHG from HCN molecule at equilibrium interacted by a multicycle laser pulse, and the dynamical interference minimum in HHG from N$_2$O molecule~\cite{Fu:oe23} has been established using TDDFT approach. In the present work, first, we utilize a much simpler theory, a singe-active-electron model with the predicted soft-Coulomb-type potential established in the previous section, to detect multiorbital effects in the HHG spectra, which are then compared with other works. Besides, we also find lower-lying orbital effects in the time-frequency spectrograms of HHG, which are more transparent when interpreting the effects. More specifically, we discover a signature of \mbox{HOMO-1} in the HHG double-plateau structure, which can be observed when using a few-cycle laser pulse.

Moreover, as a second application, we consider multiorbital effects on the dynamic process of HCN. These effects were mostly reported for molecules with nuclei fixed at their equilibrium structure~\cite{McFarland:Sci08,Smirnova:Nat09, Li:Sci08,Long:jpb23,Yun:prl15,Shu:prl22,Camper:prl23,Fu:oe23,Chu:pra23,Koushki:jmm23}. We study the variation of HHG induced by the evolution of deep-lying orbitals during the movement of the H nucleus in the H-CN stretching process. To this end, the H-C distance changes within $[2.93,~1.49]$~a.u., corresponding to $x_\mathrm{H}$ in the range of $[-3.88,~-2.44]$~a.u.. The HHG spectra are achieved numerically by solving the 2D TDSE using the split operator algorithm~\cite{Feit:JCP82}.

%%--------------------------------------------------------------------------------------------------%%
%% 4.1 Imprints of HOMO-1 in HHG from HCN molecule at equilibrium
%%--------------------------------------------------------------------------------------------------%%

\subsection{Imprints of HOMO-1 in HHG spectra from HCN molecule}

%--------------------------------------------------------------------------------------------------%%
%% FIG 4: HHG
%%--------------------------------------------------------------------------------------------------%%
\begin{figure*}[ht!]
	\begin{center}
		\includegraphics[width=0.85\linewidth]{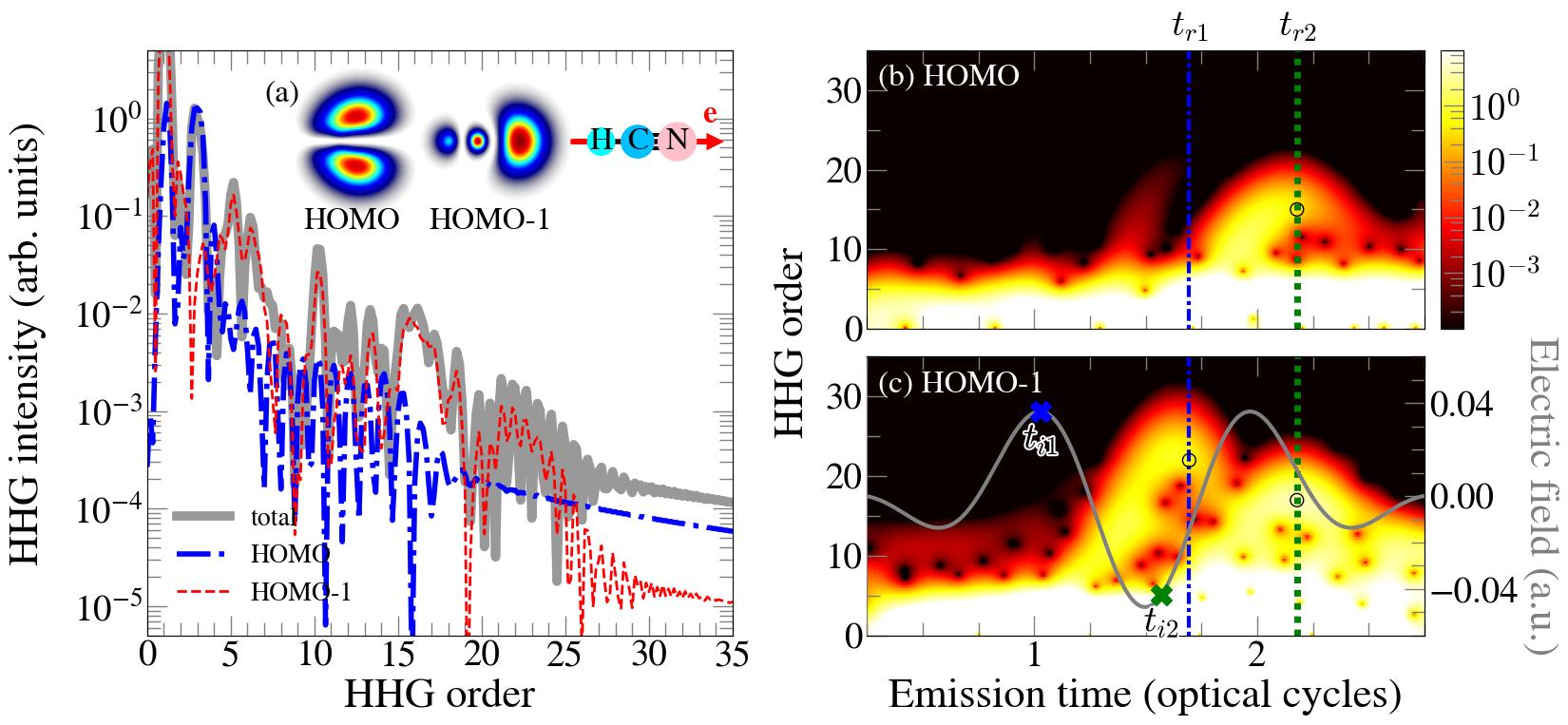}
	\end{center}
	\caption{(a) Total HHG (solid grey) from a HCN molecule and its contributions from separated HOMO (dashed-dotted blue) and HOMO-1 (dashed red) presented in spectral domain (a) and time-frequency domain [(b)-(c)]. The inset of Fig.~(a) visualizes HOMO and HOMO-1. In Fig.~(c), the gray curve presents a three-cycle sin-squared laser pulse with an intensity of $0.8\times 10^{14}$~W/cm$^2$ and wavelength of 800~nm. The laser is irradiated with the orientation angle of $0^\circ$ [see inset of Fig.~(a)]. The vertical lines in Figs.~(b) and (c) indicate the emission instants $t_{r1}$ and $t_{r2}$ of emission bursts. The corresponding ionization instants $t_{i1}$ and $t_{i2}$ are noted by blue and green crosses, respectively.}
\label{fig:HHG}
\end{figure*}
%%------------------------------------------------------------------------------------------------

In this subsection, we investigate the imprints of molecular orbitals in HHG emitted from the HCN molecule in equilibrium. Figure~\ref{fig:HHG}~(a) displays the simulation of HHG emitted from HOMO and HOMO-1 separately and their total HHG, where the electron occupation in each orbital weights the contributions of HOMO and HOMO-1. The used sine-squared envelope pulse has the form of \mbox{$\mathbf{E} = \mathbf{e} E_0 \sin^2 (\pi t/ \tau) \cos (\omega_0 t)$} where $\mathbf{e}$ is a unit vector parallel to the molecule axis and orients from the H-end to the N-end. $E_0$, $\omega_0$, and $\tau$ are the peak amplitude, carrier frequency, and pulse duration. We exemplify a three-cycle pulse with an intensity of $0.8\times 10^{14}$~W/cm$^2$ ($E_0 = 0.0424$~a.u.) and wavelength of 800~nm ($\omega_0 = 0.057$~a.u.). Here, we are interested in the contribution of HOMO and HOMO-1 only since their energies ($-0.50$~a.u. vs. $-0.58$~a.u.) have a small gap ($0.08$~a.u.), comparable with the laser frequency ($0.057$~a.u.). Moreover, their vertical ionization potentials to the two lowest states of the ion are similar~\cite{Chu:pra23}. The HOMO-2 energy ($-0.82$~a.u.) is too low, making it irrelevant in the HHG emission, as confirmed by our initial calculations (not shown).

The figures show the imprints of HOMO-1 in two features of the total HHG: \textit{intensity} and \textit{cutoff}. Regarding the \textit{intensity}, it exhibits a competition between HOMO and HOMO-1, with HOMO-1 slightly dominating HOMO for harmonics below H14. However, after this harmonic order, the emission from HOMO-1 completely surpasses the one from HOMO by about $1$ to $2$ orders of magnitude. Here, H14 denotes the 14th harmonic order. This HHG intensity behavior obtained using our ML SAE potential is consistent with that reported in the previous studies~\cite{Chu:pra23,Koushki:jmm23} when HCN interacts with a multicycle laser pulse, solved by the TDDFT method. This confirmation proves the reliability of the SAE-POT-ML model in constructing an SAE potential and reproducing molecular features. 

Regarding the second feature of HHG, i. e. the \textit{cutoff}, Fig.~\ref{fig:HHG}~(a) reveals that HHG emitted from HOMO ends much earlier than that from HOMO-1. The cutoff is at H15 for HHG from HOMO, while at H22 for HHG from HOMO-1. It should be noted that the cutoffs obtained above cannot be entirely explained by the difference in the ionization potentials of the two orbitals, as they were done for N$_2$~\cite{McFarland:Sci08} and CO$_2$~\cite{Smirnova:Nat09} molecules. Indeed, according to the semiclassical model $I_\mathrm{P} + 3.17 U_\mathrm{P}$ (where $I_\mathrm{P}$ and $U_\mathrm{P}$ are ionization potential and ponderomotive energy), the cutoffs for HOMO and HOMO-1 of the HCN molecule are respectively H19 and H20. As ones can see, these predictions agree with the TDSE simulation for the HOMO-1 only (H20 vs. H22) but fail for the HOMO (H19 vs. H15). Therefore, we will explain the TDSE results by exploring the time-frequency profile of HHG as follows.

To comprehend the signature of HOMO-1 in HHG, we look into the time-frequency (TF) profiles of HHGs from HOMO and HOMO-1 calculated by the Gabor transform~\cite{Tong:pra00}. The illustration in Fig.~\ref{fig:HHG}~(b) and (c) indicates two distinct profile structures for HOMO and HOMO-1. Particularly, Panel (c) shows that the interaction of HOMO-1 with a three-cycle laser pulse results in two attosecond bursts emitted around two instants, $t_{r1} = 1.69~T_0$ and $t_{r2} = 2.18~T_0$ ($T_0$ is an optical cycle). The HHG spectrum from HOMO-1, synthesized from these two bursts, must have the two-plateau structure with corresponding cutoffs at H17 and H22, the two brightest points in the spectrogram (c) marked by circles. Meanwhile, the HOMO's TF profile shown in Fig.~\ref{fig:HHG}~(b) mainly contains only the later attosecond burst (at the instant $t_{r2} = 2.18~T_0$). The brightest point indicates the HHG cutoff at H15. These predicted cutoff positions for both HOMO and HOMO-1 are well consistent with the cutoffs of HHG spectra shown in Fig.~\ref{fig:HHG}~(a). Moreover, the vanishing of the first burst in the TF profile of HOMO clarifies the predominant HHG intensity of HOMO-1 over HOMO's one for harmonics higher H14, i.e., for the second plateau.  

%%--------------------------------------------------------------------------------------------------%%
%% FIG 5: HHG-stretched
%%--------------------------------------------------------------------------------------------------%%

\begin{figure*}[ht]
	\begin{center}
            \includegraphics[width=0.85\linewidth]{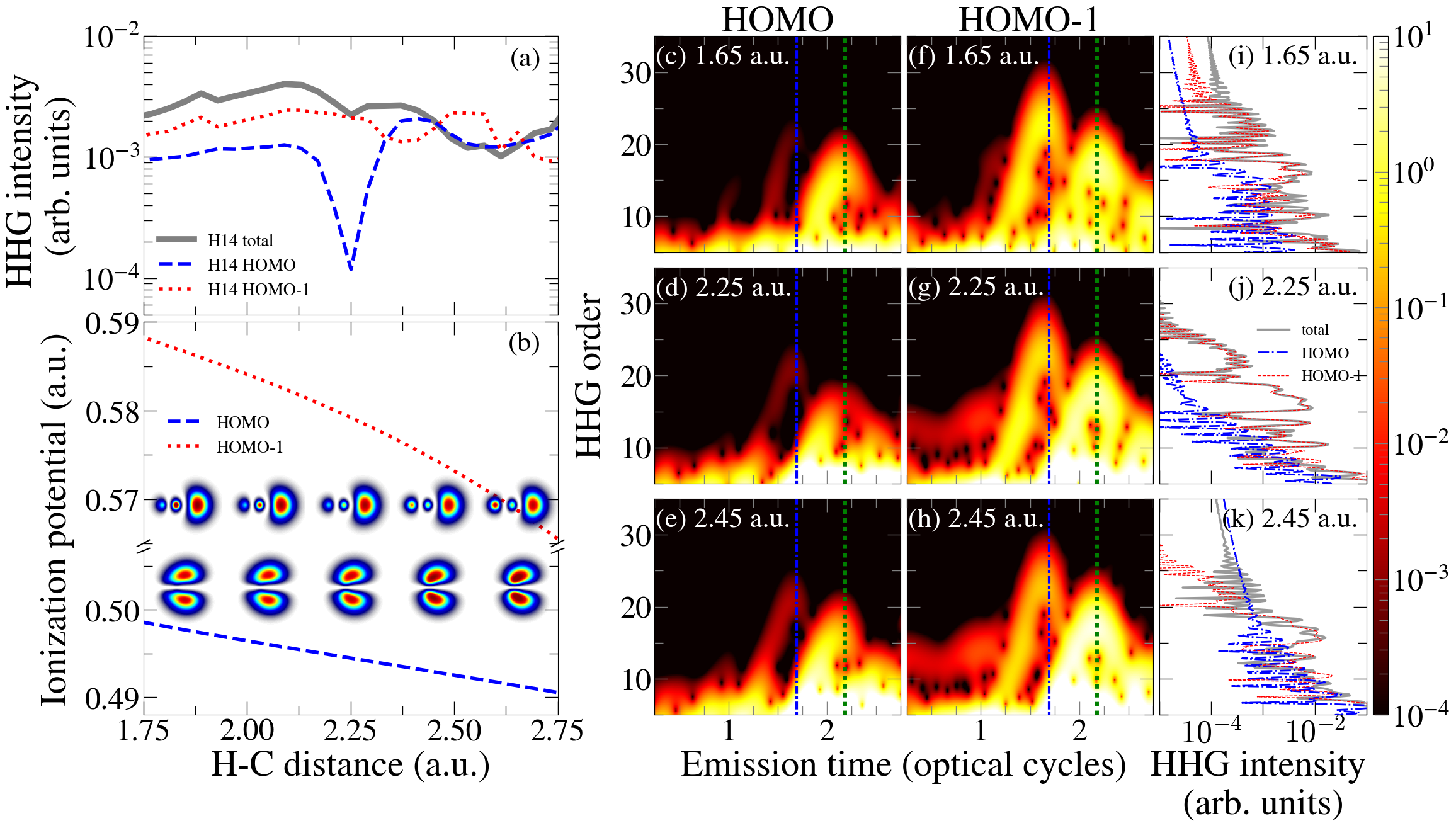}
	\end{center}
	\caption{Dependence of (a) the total and contributed harmonic intensity of order H14; and (b) ionization potential and the orbital evolution of HOMO and HOMO-1 as a function of H-C distance. Time-frequency profiles for HOMO [(c)-(e)] and HOMO-1 [(f)-(h)], and their HHG spectra [(i)-(k)] when H-C distances are 1.65~a.u., 2.25~a.u., and 2.45~a.u. The laser pulse has the same parameters as in Fig.~\ref{fig:HHG}. With increasing H-C distance, the plateau structure and smoothness of HHG changes due to the rearrangement of electron distribution. Here, the orbital redistribution can be visualized by the color change.}
\label{fig:HHG-stretched}
\end{figure*}

%%--------------------------------------------------------------------------------------------------%%

We continue to elucidate the deeper underlying of the TF profiles mentioned above by considering the correlation of the orbital spatial distribution and the ionization leading to the emission of the two attosecond bursts. According to the classical simulation of propagation of free electrons in a laser electric field, the first attosecond burst is caused by the electron ionization at instant $t_{i1} = 1.03~T_0$, denoted by the blue cross in Fig.~\ref{fig:HHG}~(c), when the electric field is considerably weak. Similarly, the second burst originates from the ionization at $t_{i2} = 1.57~T_0$ (green cross) with a more enhanced electric field. On the other hand, the HOMO-1 is the $\sigma$ orbital whose electron density is favorable along the molecule axis, while the HOMO with $\pi$ form suppresses electron density in this direction, see the inset of Fig.~\ref{fig:HHG}~(a)]. Therefore, the HOMO can be ionized when the electric field is strong enough, i.e., at $t_{i2}$, while the HOMO-1 can easily be ionized even with a weak field. As a result, the HOMO emits only the second attosecond burst, while the HOMO-1 generates both. Moreover, for HOMO-1, the N-side concentrates more electrons than the H-side. Thus, the ionization from the N-side (i.e., at $t_{i2}$) is responsible for a more enhanced emission burst (at $t_{r2}$) than that formed by the ionization from the H-side (at $t_{i1}$). Consequently, the HHG emitted from HOMO-1 has a double plateau structure. 

%%--------------------------------------------------------------------------------------------------%%
%%--------------------------------------------------------------------------------------------------%%

%%--------------------------------------------------------------------------------------------------%%
%% 4.2 Variation of plateau structure in HHG from HCN molecule during the H-CN stretching
%%--------------------------------------------------------------------------------------------------%%

\subsection{Variation of plateau structure in HHG from HCN molecule during the H-CN stretching}

We continue to investigate the HHG behavior from HCN during the H-CN stretching and find that the (i)~\textit{contribution of HOMO-1 vs. HOMO}, (ii)~\textit{plateau structure}, and (iii)~\textit{smoothness} of HHG spectra are changed with varying H-C distance, as shown in Figs.~\ref{fig:HHG-stretched}~(i), (j), and (k) for three distances: 1.65, 2.25, and 2.45 a.u.. 
The change is mainly in the second plateau, originally emitted by HOMO-1. The contribution of HOMO-1 is substantial for small H-C distances but then rapidly weakens for more considerable distances. Meanwhile, the first plateau is contributed by both HOMO and HOMO-1, where the HOMO-1 slightly defeats the HOMO. We exemplify the first plateau change by considering the harmonic H14 shown in Fig.~\ref{fig:HHG-stretched}~(a), where the HHG intensity is almost unchanged during the movement of the H-atom. Lastly, Figs.~\ref{fig:HHG-stretched}~(i), (j), and (k) also show the HHG spectra, which are more and more smooth with increasing the H-C distance, especially in the first plateau.

The origin of the changes mentioned above can be understood through the time-frequency profiles of HHG emitted from HOMO and HOMO-1 with different H-C distances, as illustrated in Figs.~\ref{fig:HHG-stretched}~(c)-(f), (d)-(g), and (e)-(h). Indeed, Figs.~\ref{fig:HHG-stretched}~(c), (d), and (e) indicate that the HHG from HOMO is contributed mainly by the emission of the attosecond burst at $t_{r2}$ (the second burst), where its intensity is considerably small for the intermediate H-C distance, say about 2.25~a.u. in Panel (d). It is consistent with the H14 harmonic intensity of HOMO shown in Fig.~\ref{fig:HHG-stretched}~(a). At a large H-C distance, 2.45 a.u. in Panel (e), the first attosecond burst appears at $t_{r1}$, although it is still relatively weak that the second plateau does not appear in the HHG from HOMO. 
Meanwhile, Figs.~\ref{fig:HHG-stretched}~(f), (g), and (h) for HOMO-1 show two attosecond bursts at $t_{r1}$ and $t_{r2}$ for three H-C distances, meaning the HHG from HOMO-1 is contributed from both emission bursts during the H-C stretching process. It is consistent with the two-plateau structure of HHG from HOMO-1. However, the relative intensities of the two bursts vary with changing H-C distance. The second burst slightly enhances compared to the first for the small and intermediate H-C distances, but it strongly dominates in the large distance in Panel (h), explaining the vanishing of the second HHG plateau of HOMO-1 in Fig.~\ref{fig:HHG-stretched}~(k). 
Moreover, the time-frequency profiles can explain the smoothness of the HHG. Indeed, since the HHG is formed mainly by the interference of two branches of the second emission burst (corresponding to the long and short trajectories) at a large H-C distance, the spectral HHG is smoother for this case. Meanwhile, for a small H-C distance, the interference of four branches from both bursts causes a more complex structure in spectral HHG.   

Now, to explain the underlying physics of the above discussions, we pay attention to the change of the ionization potential and the rearrangement of the orbital's spatial electron distribution during the H-C distance stretching, as demonstrated in Fig.~\ref{fig:HHG-stretched}~(b). It shows that the HOMO's ionization potential decreases as the H-C distance increases, making it easier for electrons to escape. Simultaneously, electrons transfer from the N-side at a small H-C distance into the H-side at a large H-C distance, enhancing the first emission burst. At an intermediate distance, electrons are trapped in the middle region, reducing electron distribution both on the H-side and N-side, causing the decrease of emission in both attosecond bursts and, subsequently, leading to the drop in the HHG intensity. 
For HOMO-1, when the H-nucleus is very close to the C-nucleus at a small H-C distance, their joint electron distribution is still significant; thus, the first emission burst (from the HC side) is almost as strong as the second one (from the N side). At a large H-C distance, i.e., the H-nucleus is far away from the CN part, the electron distribution on the H-side is reduced compared to that in the previous case of small H-C distance, lowering the intensity of the first attosecond burst. In this case, although the electron distribution of the N-side is reduced, the ionization potential is also reduced as well. Together, these two factors make the emission of the second attosecond burst strong.        

In short, the fascinating variation of HHG spectra during H-CN stretching comes from the electron redistribution of orbitals. We have successfully explored the HOMO-1 effects by using the predicted ML molecular potential because it gives not only precise energies for HOMO and HOMO-1 but also their orbital forms and orbital electric dipoles during the movement of the H atom.    
 
%%--------------------------------------------------------------------------------------------------%%
%%--------------------------------------------------------------------------------------------------%%

%%=====================================================================================================================================================
%% SECTION 5: Conclusions
%%=====================================================================================================================================================

\section{Conclusions and Outlook}

In summary, we have built an ML model as a fast, effective, and reliable tool for constructing the soft-Coulomb-type SAE potential of a polar molecule that can describe the molecular features of not only HOMO but also deep-lying orbitals. We have then successfully applied the ML approach for a triatomic polar HCN molecule whose SAE potential containing six independent parameters is constructed from the molecular feature set comprising orbital symmetries, energies, and electric dipole moments of HOMO, HOMO-1, and HOMO-2. This so-called SAE-POT-ML model works well for HCN molecules at equilibrium and in the H-CN stretching process.  

The construction of the SAE-POT-ML model has been performed in two steps. First, the SAE ML model with DCNN architecture is built to predict the SAE potential in the image form from the molecular feature set. Then, the pic2para ML model is constructed using the CNN architecture to convert the potential image to the potential parameters. Transfer learning, meaning retraining the ML model of fixed HCN at equilibrium with moderate data, has been implemented to train the ML model in the case of a moving H atom. Once trained, the SAE-POT-ML model can quickly predict six parameters of the SAE potential in less than a minute. The constructed SAE potential gives exact orbital symmetries, remarkably accurate energies, and electric dipole moments. The fast and reliable performance of the SAE-POT-ML model is especially meaningful in repeatedly yielding molecular potentials during molecular dynamics.

Also, applications of the constructed SAE potential were given to explore the multiorbital effects in high-order harmonic generation from the HCN molecule. We studied the imprints of HOMO-1 in HHG emitted when a three-cycle laser pulse irradiates along the molecular axis of the HCN molecule during the H atom's movement in the \mbox{H-CN} stretching process. We have revealed the HOMO-1’s effects in both the time-frequency and spectral domains of HHG. In the time-frequency spectrogram, HOMO-1 generates a two-attosecond-burst pattern instead of the one-burst of HOMO. As a result, in the spectral domain, \mbox{HOMO-1} causes the emergence of the second plateau, leading to the corresponding enhancement of HHG intensity and high cutoff energy compared to HOMO’s HHG. However, with increasing the H-C distance, the second plateau is suppressed compared to the first due to the modified orbital ionization potentials and the redistribution of the orbital electron density. Our findings supply a comprehensive HOMO-1 effect on HHG throughout a molecular dynamic process. 

Although the application of the SAE-POT-ML is restricted to the H-CN stretching process in this work, it can straightforwardly be extended to various dynamic processes, such as the isomerism of HCN to HNC. Moreover, the approach to build the ML model for SAE potentials in this study is general to be applied to other two- or three-atom molecules. Further research is necessary to determine how this approach can be applied to more complex molecules.

%%=====================================================================================================================================================
%% SECTION 6: Acknowledgements
%%=====================================================================================================================================================
\section*{Data and code availability}
The data and codes that support the findings of this study are available from the corresponding author upon reasonable request.

\section*{Author contributions}

N.-L.P. and V.-H.L. conceived and supervised the project. D.D.H.-T., K.T., and N.-L.P. designed machine learning methods. D.D.H.-T., D.-A.T., Q.-H.T., and N.-L.P. provided numerical methods. D.D.H.-T., K.T., D.-A.T., N.-L.P. generated data and performed analysis. N.-H. P. supported validating the results. D.D.H.-T., N.-L.P., and V.-H.L. wrote the first draft. All authors contributed to the data interpretation and finalizing the manuscript.

\section*{Conflicts of interest}
There are no conflicts to declare.
\section*{Acknowledgements}
We are grateful to Kim-Ngan H. Nguyen, Cam-Tu Le, DinhDuy Vu, and Ngoc-Ty Nguyen for their insightful discussions and and useful suggestions. This work was funded by Vingroup and supported by Vingroup Innovation Foundation (VINIF) under project code VINIF.2021.DA00031. The calculations were executed by the high-performance computing cluster at Ho Chi Minh City University of Education, Vietnam.

\section*{Supplementary Material}
The Supplementary Material contains:
\begin{itemize}
  \item Supplementary A: Construction of SAE ML model
  
  In this Supplementary, we detail the construction and training preSAE ML model (for HCN molecule fixed at equilibrium position) and SAE ML model (for HCN during the H-CN stretching process), which together form the first step in the two-step SAE-POT-ML model.

  \item Supplementary B: Construction of pic2para model

   This Supplementary presents the second step of the two-step SAE-POT-ML model: developing and training 
   the pic2para model. This model converts the 2D image of SAE potential into its parameters as indicated from Eq. (\ref{eq:V_SAE}).
   
\end{itemize}

%%=====================================================================================================================================================
%% References
%%=====================================================================================================================================================

\bibliography{MyBib}
\bibliographystyle{IEEEtran}

\end{document}

\preprint{AIP/123-QED}

\title
[Supplementary Material for Machine-Learning-Based Construction of Molecular Potential ...
% and Its Application in Exploring the Deep-Lying-Orbital Effect in High-Order Harmonic Generation
]
{Supplementary Material for \\
Machine-Learning-Based Construction of Molecular Potential and \\
Its Application in Exploring the Deep-Lying-Orbital Effect in \\
High-Order Harmonic Generation
}

\author{Duong D. Hoang-Trong}
    \affiliation{Computational Physics Key Laboratory K002, Department of Physics, Ho Chi Minh City University of Education, 280 An Duong Vuong Street, Ward 4, District 5, Ho Chi Minh City 72711, Vietnam}

\author{Khang Tran}
    \affiliation{Department of Informatics, New Jersey Institute of Technology, Newark, NJ 07102, USA}

\author{Doan-An Trieu}
    \affiliation{Computational Physics Key Laboratory K002, Department of Physics, Ho Chi Minh City University of Education, 280 An Duong Vuong Street, Ward 4, District 5, Ho Chi Minh City 72711, Vietnam}

\author{Quan-Hao Truong}
    \affiliation{Computational Physics Key Laboratory K002, Department of Physics, Ho Chi Minh City University of Education, 280 An Duong Vuong Street, Ward 4, District 5, Ho Chi Minh City 72711, Vietnam}

\author{Ngoc-Hung Phan}
    \affiliation{Computational Physics Key Laboratory K002, Department of Physics, Ho Chi Minh City University of Education, 280 An Duong Vuong Street, Ward 4, District 5, Ho Chi Minh City 72711, Vietnam}

\author{Ngoc-Loan Phan} 
    \affiliation{Computational Physics Key Laboratory K002, Department of Physics, Ho Chi Minh City University of Education, 280 An Duong Vuong Street, Ward 4, District 5, Ho Chi Minh City 72711, Vietnam}

\author{Van-Hoang Le}
    % \email{hoanglv@hcmue.edu.vn.}  
    \affiliation{Computational Physics Key Laboratory K002, Department of Physics, Ho Chi Minh City University of Education, 280 An Duong Vuong Street, Ward 4, District 5, Ho Chi Minh City 72711, Vietnam}
    
\date{\today}% It is always \today, today,

\maketitle
\onecolumngrid
% The Supplementary Material contains:
% \begin{itemize}
%   \item Supplementary A: Construction of SAE ML model
  
%   In this Supplementary, we detail the construction and training preSAE ML model (for HCN molecule fixed at equilibrium position) and SAE ML model (for HCN during the H-CN stretching process), which together form the first step in the two-step SAE-POT-ML model.

%   \item Supplementary B: Construction of pic2para model

%    This Supplementary presents the second step of the two-step SAE-POT-ML model: developing and training 
%    the pic2para model. This model converts the 2D image of SAE potential into its parameters as indicated from Eq. (\ref{MyMain-eq:V_SAE}) in the Manuscript.
   
% \end{itemize}

\appendix

%%=====================================================================================================================================================
%% SUP A: Construction of SAE ML model
%%=====================================================================================================================================================

\section{Supplementary A: Construction of SAE ML model}
\label{Sup1}
\renewcommand{\thefigure}{A.\arabic{figure}}
\setcounter{figure}{0}
\renewcommand{\thetable}{A.\arabic{table}}
\setcounter{table}{0} 

In this Supplementary, we provide details of how to build and train the SAE ML potential model, which is the first step in the two-step SAE-POT-ML model. The SAE ML model predicts the HCN potential image with correct asymptotic behavior from the molecular feature set (energies, electric dipoles, and symmetries of HOMO, HOMO-1, HOMO-2). Our ambition is to generate the SAE ML potentials at any point during HCN dynamic processes. For this aim, we first build the preSAE ML model for HCN molecule fixed at equilibrium position, then take advantage of transfer learning to construct the SAE ML model for HCN during the H-CN stretching process with much smaller data demand based on the pre-trained preSAE ML model. Before discussing the ML model, we present data generation in detail.     

%%--------------------------------------------------------------------------------------------------
%% A.1. Data generation and feature engineering
%%--------------------------------------------------------------------------------------------------

\hspace{0.3 cm}

\noindent \textit{A.1. Data generation and feature engineering} 

\hspace{0.1 cm}

For training the model, the datasets are generated by solving TISE, where each data point is made of the potential parameter set (six potential parameters) and molecular feature set. Here, the energy of orbital $i$ is defined as $E_i = \bra{\psi_i (x,y)} \hat{\mathrm{H}}_0 \ket {\psi_i (x,y)}$ where $\hat{\mathrm{H}}_0$ is the Hamiltonian of the molecule HCN. The orbital dipole \mbox{$d_i = -\bra{\psi_i (x,y)} x \ket {\psi_i (x,y)} $} is aligned along the $x-$axis. The orbital symmetries $s$ are estimated via $\braket{\psi_i (x,y)|\psi_i (x,-y)}$ whose values are $\approx +1$ and $\approx - 1$ for $\sigma$ and $\pi$ orbitals, respectively. The dataset of 30,000 data points for the case of HCN is fixed at equilibrium, and 100.000 data points for H atom moving in the range of $[-4.0, -2.4]$~a.u.

We perform feature engineering from the raw data to transform them into features. To this end, we divide the molecular feature set into two parts: one contains continuous data -- the orbital energies, dipoles, and position of H atom; and the other comprises binary data -- orbital symmetries. The first part is treated by the Polynomial Feature Transformation (PFT) with the fourth-degree polynomial and Principal Component Analysis (PCA). These techniques can retain the most information while transforming the original data into new dimensions that improve its interpretability~\cite{Zhu:IMU2019}. The second part is rescaled with binary values as 1 for $\sigma$ orbital and 0 for $\pi$ one. Then, the second data part is concatenated with the first one to create a fully new dataset with 128 features.  

Concerning the potential image, two notices should be noted. First, since the HCN molecule is linear, its potential image is axially symmetric around $y = 0$. To reduce the computational cost, we train the model to predict half of the potential image only, and then reflect it over $y = 0$ to get the full one. Second, since the most important information of the potential image is concentrated around the center of the image while its surroundings contain less information, we train two separate models to improve the model's accuracy. The first one predicts the center half-image with the size of $x \times y = [-4.6,~3.4) \times [0,~2.0)$~a.u. and the latter predicts the entire half-image with $x \times y =[-20,~20) \times [0,~20)$~a.u.. The spatial grid is 0.2~a.u. After training, we paste the predicted center half-image over the predicted entire one to get the final half-image.

%%--------------------------------------------------------------------------------------------------%%
%%--------------------------------------------------------------------------------------------------%%

%%--------------------------------------------------------------------------------------------------%%
%% A.2. PreSAE ML model for HCN molecule at equilibrium
%%--------------------------------------------------------------------------------------------------%%

\hspace{0.3 cm}

\noindent \textit{A.2. PreSAE ML model for HCN molecule at equilibrium}

\hspace{0.1 cm}

The pre-SAE model predicts the potential image of HCN molecule whose atoms are fixed at equilibrium positions. It consists of  preSAE-0 ML and preSAE-1 ML models for predicting the center and entire potential half-images from the molecular feature set. 

%%--------------------------------------------------------------------------------------------------%%
%% FIG A1: model1
%%--------------------------------------------------------------------------------------------------%%
\begin{figure}[htb]
	\begin{center}
		\includegraphics[width=0.7\linewidth]{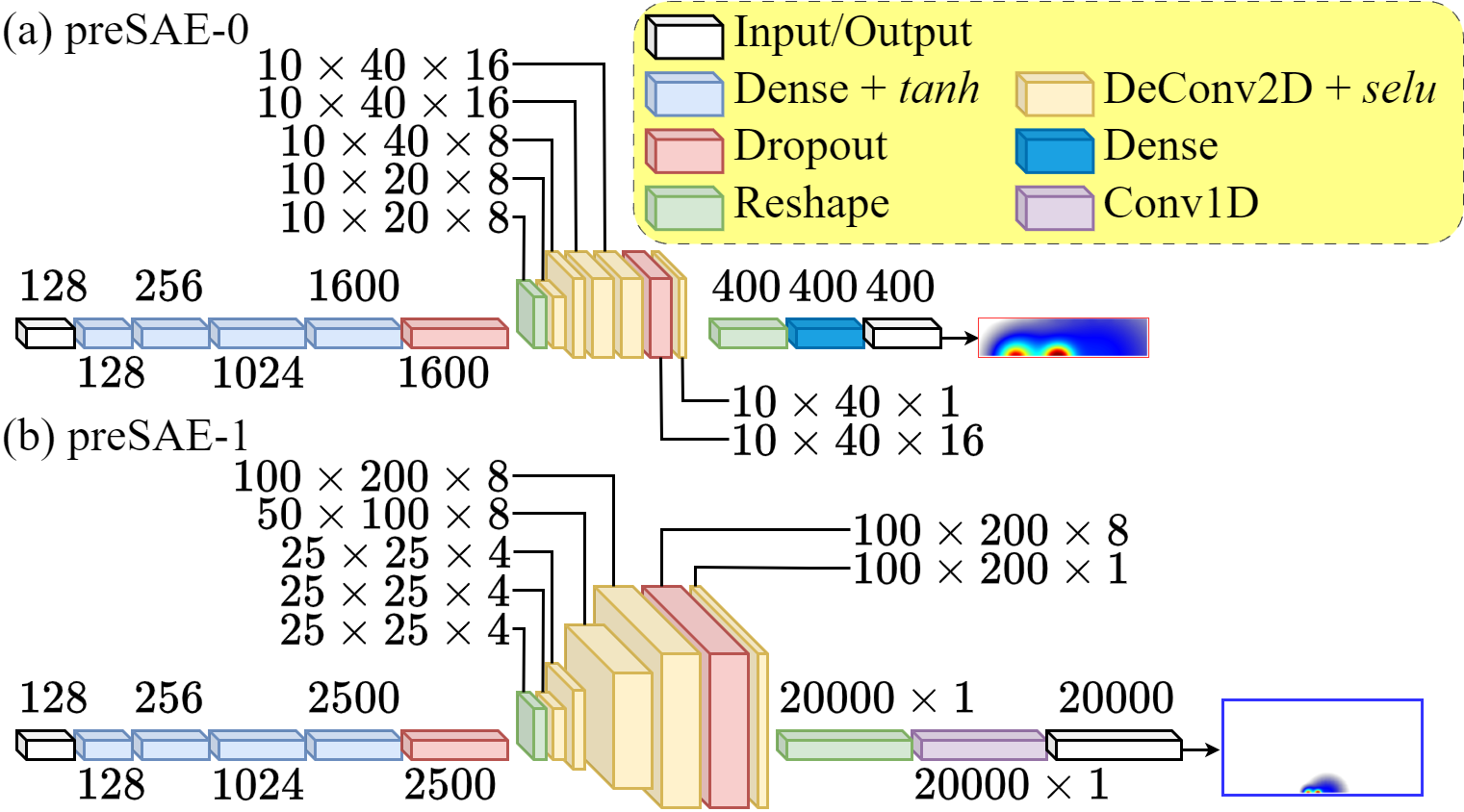}
	\end{center}
	\caption{Architectures of preSAE-0 (a) and preSAE-1 (b) models to predict the center and the entire half-images, respectively. Each model contains a fully connected, DCNN, and smoothening blocks. The numbers show the output shape of each layer. The legends describe the type of each layer and its activation function.}
\label{fig:model1}
\end{figure}
%%--------------------------------------------------------------------------------------------------%%
%%--------------------------------------------------------------------------------------------------%%

\textit{Architecture --} Figure \ref{fig:model1} depicts the DCNN architecture for preSAE-0 ML and preSAE-1 ML models to predict the center and entire potential half-images. Each model is constructed from three main blocks: the fully connected block (the first six layers), the deconvolutional block (the following seven layers), and the smoothening block (the last three). In the first block, the input is passed through the fully connected layers using activation function \textit{tanh} to extract the data features, making generating SAE potential images more efficient. The dropout layer is employed to avoid overfitting. The second block is started by the reshape layer to convert the 1D array into a 2D array. Then, the deconvolutional neural network constructs and enlarges the SAE images at each 2D-deconvolutional layer using different deconvolution filters. Here, the \textit{selu} activation function is used between each layer, effectively generating images~\cite{Xu2018}. Finally, the last block aims to sharpen and smoothen the generated image by the last three layers: one reshaped layer, one fully connected layer in the preSAE-0 ML model, or one 1D-convolutional layer in preSAE-1 ML, and the output. The images of the potential generated at the last layer of the model are called predicted images.

%%--------------------------------------------------------------------------------------------------%%
%% FIG A2: process1
%%--------------------------------------------------------------------------------------------------%%
\begin{figure} [htb]
	\begin{center}
		\includegraphics[width=0.75\linewidth]{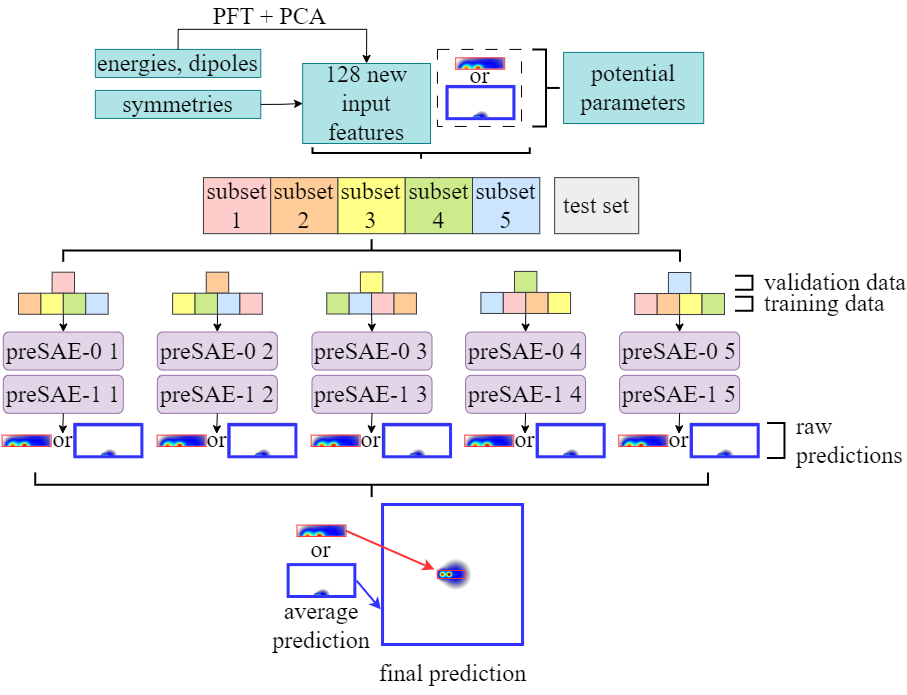}
	\end{center}
	\caption{Training process of preSAE-0 and preSAE-1 ML models with five-fold cross-validation technique.}
\label{fig:process1}
\end{figure}
%%--------------------------------------------------------------------------------------------------%%
%%--------------------------------------------------------------------------------------------------%%

{\textit {Training --}}The main training process is exhibited in Fig.\ref{fig:process1}. We divided the dataset into the training set and test set, which are made up respectively 80~\% and 20~\% of the total. Then, to improve training efficiency with the moderate training data set, we implement the five-fold cross-validation technique. The training set is randomly partitioned into five distinct subsets called folds. The preSAE ML model is then trained and evaluated five times; each time, one fold is selected for model validation while the other four folds are implemented for training. As a result, we have five separately predicted SAE images, which are then averaged to obtain the final SAE image. This technique, known as ensemble learning, increases model performance while avoiding overfitting~\cite{Aurelien:19}.

The learning process minimizes the difference between the true and predicted outputs. In each iterative during the training process, the difference is determined by the loss function
%%--------------------------------------------------------------------------------------------------%%
%% EQ A1: loss_func1  New
%%--------------------------------------------------------------------------------------------------%%
\begin{equation}
\label{eq:loss_func1}
L=\frac{1}{N}\displaystyle\sum\limits_{j=1}^N 
{
\scaleleftright[1.ex]
{(}{
\sqrt{\frac{1}{n} \displaystyle\sum\limits_{i=1}^n |\hat{z}_i [j]-z_i [j]|^2} +
\frac{1}{n} \displaystyle\sum\limits_{i=1}^n |\hat{z}_i [j]-z_i [j]|
}{)}
}
\end{equation}
%%--------------------------------------------------------------------------------------------------%%
%%--------------------------------------------------------------------------------------------------%%
where $\hat{z}_i [j]$ and $z_i [j]$ are the predicted and true values of pixel $i$ in image $j$, while $N$ and $n$ are, respectively, the total number of images and of pixels in each image. The first term in Eq.~(\ref{eq:loss_func1}) is known as the Root Mean Square Error (RMSE), and the second one is the Mean Absolute Error (MAE). Although RMSE is appropriate to predict SAE potential images with small errors, it may blunt and unsmooth the image. Therefore, we add the MAE to balance the accuracy and smoothness of the predicted outcome~\cite{Zhao:tci17}. To optimize the process minimizing loss function, the optimizer Adam is implemented to update the model, i.e., the weights (in fully connected layers) and filters (in deconvolutional layers) with the learning rate $l_r=0.001$ for each iteration, which is referred to as an epoch. To avoid overfitting, the process stops if the loss function is not improved after at least 50 epochs~\cite{Aurelien:19}. The batch size is 512. The training phase indicates that the loss function rapidly decreases and converges after about 20-30 epochs (not shown).

{\textit {Validating and testing --}} After training the model, we validate and test the preSAE-0 and preSAE-1 ML models with the Mean Absolute Percentage Error (MAPE) metric, which is defined for each fold as
%%--------------------------------------------------------------------------------------------------%%
%% EQ A2: metric
%%--------------------------------------------------------------------------------------------------%%
\begin{equation}
\mathrm{MAPE}=\frac{1}{N}\displaystyle\sum\limits_{j=1}^N 
{\left(
\displaystyle\sum\limits_{i=1}^n \frac{1}{n} \frac{|\hat{z}_i[j]-z_i[j]|}{|z_i[j]|} 
\right)}.
\label{eq:metric}
\end{equation}
%%--------------------------------------------------------------------------------------------------%%
%%--------------------------------------------------------------------------------------------------%%
The same formula takes the final MAPE, but the predicted SAE potential image $\hat{z}_i[j]$ of each fold is now replaced by the average one over all five folds, i.e., $\overline{\hat{z}_i[j]}$. Our results show that the MAPE in the test phase of separated folds varies in $[1.98,~	2.07]\%$ and $[2.80,~6.44]\%$ for preSAE-0 and preSAE-1 models, respectively. The values of each fold are not significantly different, indicating that the models are well-designed and robust. On the other hand, the errors of these folds compensate each other, lowering the final MAPE, which are $1.66\%$ and $2.08\%$ for preSAE-0 and preSAE-1 models. These low errors demonstrate the potency of ensemble learning. Moreover, the comparison MAPEs of the independent validation and test datasets show similar errors in each fold, manifesting the good performance of our preSAE models.

%%--------------------------------------------------------------------------------------------------%%
%%--------------------------------------------------------------------------------------------------%%

%%--------------------------------------------------------------------------------------------------%%
%% A.3. SAE ML model for HCN molecule with moving H atom during H-CN stretching process
%%--------------------------------------------------------------------------------------------------%%
\hspace{0.3 cm}

\noindent \textit{A.3. SAE ML model for HCN molecule with moving H atom during H-CN stretching process}

\hspace{0.1 cm}

After constructing the reliable preSAE ML model for predicting the SAE potential image of the HCN molecule at equilibrium, we apply transfer learning to predict the potential for the HCN molecule with the moving of H atom during the H-CN stretching process. Similar to the preSAE model, the SAE ML model trains half of the 2D potential image, and then we get its reflection to get the full image. Besides, we also built the two models, named \mbox{SAE-0 ML} and SAE-1 ML models, to train the center and entire half-images separately. The final image is obtained in the same way as in the preSAE model.      

{\textit {Training --}} We reuse the method of the preSAE model to partition the dataset into the training, validation, and test sets; and the training set into folds. For training, we freeze the weights of the deconvolutional layer blocks as of pretrained preSAE models and retrain the rest of the architecture. The hyperparameter settings (optimizer, learning rate, etc.) are left unchanged. The loss function values $L$ rapidly converge after 20-30 epochs.  

{\textit {Validating and testing --}} To validate and test the trained SAE-0 and SAE-1 ML models, we calculate MAPEs. The final MAPEs for these models are respectively 4.49~\% and 8.44~\%, not much larger than those of the pretrained models, demonstrating the effectiveness of transfer learning with a much smaller dataset.      

%%--------------------------------------------------------------------------------------------------%%
%%--------------------------------------------------------------------------------------------------%%

%%=====================================================================================================================================================
%% SUP B: Construction of pic2para model
%%=====================================================================================================================================================

\section{Supplementary B: Construction of pic2para model}
\label{Sup2}
\renewcommand{\thefigure}{B.\arabic{figure}}
\setcounter{figure}{0}
\renewcommand{\thetable}{B.\arabic{table}}
\setcounter{table}{0}

The SAE potential in the analytical form is more convenient than the image one in solving diverse chemical and physical interactions. As indicated from Eq.~(\ref{MyMain-eq:V_SAE}) in the Manuscript, there are six parameters characterizing the SAE potential of HCN molecule, including $a_\alpha$ and $\sigma_\alpha$ with $\alpha = \mathrm{H,~C,~N}$. Therefore, in this section, we build a model to convert the 2D image of SAE potential into its parameters. Hence, we train separated ML models with identical architecture to predict each potential parameter.

{\it {Data generation --}} The dataset is easily generated using Eq. (\ref{MyMain-eq:V_SAE}) in the Manuscript with the parameters in the range listed in Tab.\ref{MyMain-tab:parameters} in the Manuscript. The obtained 2D potential images are then cropped (with the same size used in the SAE ML model) to get the center image since they include the most crucial information.

%%--------------------------------------------------------------------------------------------------%%
%% FIG B1: model2
%%--------------------------------------------------------------------------------------------------%%
\begin{figure}[htb]
	\begin{center}
		\includegraphics[width=0.6\linewidth]{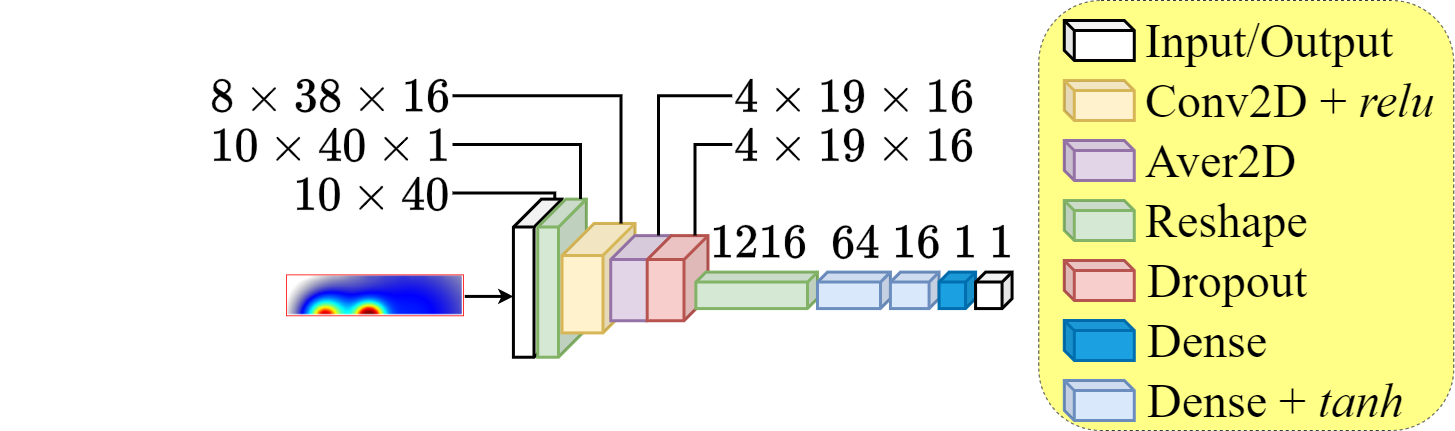}
	\end{center}
	\caption{Architectures of pic2para model to predict potential parameters from its image.}
\label{fig:model2}
\end{figure}
%%--------------------------------------------------------------------------------------------------%%
%%--------------------------------------------------------------------------------------------------%%

%%--------------------------------------------------------------------------------------------------%%
%% FIG B2: process2
%%--------------------------------------------------------------------------------------------------%%
\begin{figure} [!ht]
	\begin{center}
		\includegraphics[width=0.7\linewidth]{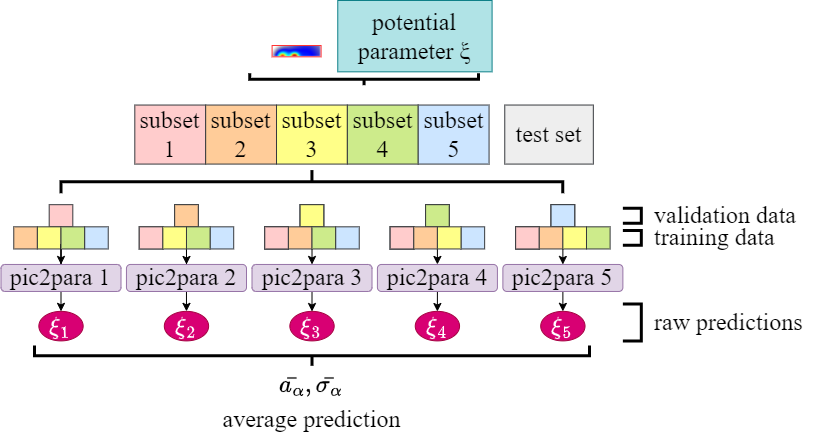}
	\end{center}
	\caption{Training process of pic2para models.}
\label{fig:process2}
\end{figure}
%%--------------------------------------------------------------------------------------------------%%
%%--------------------------------------------------------------------------------------------------%%

{\it {Architecture --}} Figure \ref{fig:model2} shows the model architecture, which contains two main blocks: the convolutional block (the first five layers) and the fully connected block (the last five layers). In the first block, the input passes through the convolutional layers with the activation function \textit{relu} to extract its features. The last 2D CNN layer flattens the images. In the next block, the fully connected layers with the activation function \textit{tanh} receive the 2D image and then determine its relationship with the parameters, which are called predicted potential parameters.    

{\it {Training --}} We also apply the five-fold crossing-validation technique of ensemble learning, as shown in Fig.~\ref{fig:process2}. The training process minimizes the lost function
%%--------------------------------------------------------------------------------------------------%%
%% EQ B2: loss_func2
%%--------------------------------------------------------------------------------------------------%%
\begin{equation}
L=\frac{1}{n} \displaystyle\sum\limits_{i=1}^n |\hat{z}_i - z_i|^2,
\label{eq:loss_func2}
\end{equation}
%%--------------------------------------------------------------------------------------------------%%
%%--------------------------------------------------------------------------------------------------%%
where $\hat{z}_i$ and $z_i$ respectively are predicted and true values of potential parameters, while $n$ is the number of data points. The Mean Square Error (MSE) loss function is commonly employed in ML models doing the same regression task (predict numbers from images) \cite{jog2013,Chernikova2019}. The weights (in fully connected layers) and the filters (in CNN block) are updated by the optimizer Adam with the learning rate $l_r=0.001$ for each epoch and stop when the loss function value is unimproved after 30 epochs to avoid overfitting \cite{Aurelien:19}. 

{\it {Validating and testing}} - We estimate the MAPE of each model training each potential parameter. The calculated MAPEs for each fold of each model are accurate and stable in both the validation and test phases. The final MAPEs of parameters are less than 4.67~\% for N and C atoms and are less than 20.33~\% for H atom, showing the robustness and reliability in extracting potential parameters from its image. 

Besides evaluating MAPE, we perform the second test by comparing the molecular feature set (predicted set) from solving TISE using the predicted SAE potential images and the one used for training the SAE ML model (true set). As discussed in the main text, the predicted molecular features are highly consistent with the input features calculated by GAUSSIAN.

% Program name: SAE-POT-ML. 

% Programming language: Python.   

% Program description: This SAE-POT-ML program constructs the soft-Coulomb-type SAE potential of the molecule HCN at equilibrium and during the H-CN stretching from molecular properties of multiple orbitals, including energies, symmetries, and dipole moments. This program is available on the Code Ocean platform.

% {\em Input:} The molecular feature sets include the orbital symmetries, energies, dipole moments of HOMO, HOMO-1, and HOMO-2, and the position of H atom (for the H-CN stretching process). Those molecular feature sets can be obtained from experiments, or estimated from chemical source codes such as GAUSSIAN, GAMESS, ORCA.

% {\em Output:} The SC SAE potential of HCN molecules in two forms - the image form (2D array), and the analytical form (Eq.~\ref{eq:V_SAE}) with six exported parameters $a_\mathrm{H}$, $\sigma_\mathrm{H}$, $a_\mathrm{C}$, $\sigma_\mathrm{C}$, $a_\mathrm{N}$, $\sigma_\mathrm{N}$.  

%%=====================================================================================================================================================
%% References
%%=====================================================================================================================================================

\bibliography{MyBib}
\bibliographystyle{IEEEtran}